\documentclass[12pt,a4paper]{article}
\usepackage[latin1]{inputenc}
\usepackage{a4wide}
\usepackage{amsfonts}
\usepackage{amssymb}
\usepackage{amsmath}
\usepackage{bbm}
\usepackage{ifpdf}
\ifpdf
\usepackage[pdftex]{graphicx}
\usepackage[pdftex,unicode,implicit]{hyperref}
\hypersetup{%
  pdftitle    = {The supersymmetric tensor hierarchy of N=1,d=4 supergravity},
  pdfkeywords = {Supersymmetry, supergravity, symmetry, gauge, },
  pdfauthor   = {Jelle Hartong, Mechthild Huebscher and  Tomas Ortin},
  pdfcreator  = {pdf\LaTeXe\ with package \flqq hyperref\frqq},
  pdfproducer = {pdf\LaTeXe\ with package \flqq hyperref\frqq},
  pdfpagemode = None,  
  pdffitwindow= true,  
  unicode     = true,
  plainpages  = true,
  colorlinks  = true,  
  citecolor   = blue,
  urlcolor    = red,
  linkcolor   = black
}
\newcommand{\hepth}[1]{arXiv:{\tt
\href{http://www.arXiv.org/abs/hep-th/#1}{hep-th/#1}}}

\newcommand{\arxiv}[1]{{\tt
\href{http://www.arXiv.org/abs/#1}{arXiv:#1}}}
\else
  \usepackage[dvips]{graphicx}
  \usepackage[unicode,implicit]{hyperref}
  \newcommand{\hepth}[1]{arXiv:{\tt hep-th/#1}}

  \newcommand{\arxiv}[1]{{\tt arXiv:#1}}
\fi
\makeatletter
\@addtoreset{equation}{section}
\makeatother

\makeindex
\begin{document}

\begin{flushright}
\small
IFT-UAM/CSIC-09-02\\
\today\\
\normalsize
\end{flushright}

\begin{center}

\vspace{.7cm}

{\LARGE {\bf The supersymmetric tensor hierarchy of\\[.5cm]
 $N=1,d=4$ supergravity}} 

\vspace{1cm}

\begin{center}

{\bf Jelle Hartong ${}^{\dagger}$, 
Mechthild H\"ubscher ${}^{\ddagger}$, and Tom\'as Ort\'{\i}n} ${}^{\ddagger}$ 

\vspace{.7cm}

${}^{\dagger}$ {\em Institute for Theoretical Physics, \\
Sidlerstrasse 5, CH-3012 Bern,
Switzerland\vskip 5pt}

{e-mail: {\tt hartong@itp.unibe.ch}}
\vskip 15pt

${}^{\ddagger}$ {\em Instituto de F\'{\i}sica Te\'orica UAM/CSIC
Facultad de Ciencias C-XVI, \\
C.U. Cantoblanco, E-28049-Madrid, Spain\vskip 5pt}

{e-mail: {\tt Mechthild.Huebscher@uam.es, Tomas.Ortin@uam.es}}

\end{center}

\vspace{1cm}

{\bf Abstract}

\begin{quotation}

  {\small 
    In this paper we construct the supersymmetric tensor hierarchy of N=1, d=4
    supergravity. We find some differences with the general bosonic
    construction of 4-dimensional gauged supergravities.

    The global symmetry group of $N=1,d=4$ supergravity consists of three
    factors: the scalar manifold isometry group, the invariance group of the
    complex vector kinetic matrix and the $U(1)$ R-symmetry group. In contrast
    to (half)-maximal supergravities, the latter two symmetries are not
    embedded into the isometry group of the scalar manifold.  We identify some
    components of the embedding tensor with Fayet-Iliopoulos terms and we find
    that supersymmetry implies that the inclusion of R-symmetry as a factor of
    the global symmetry group requires a non-trivial extension of the standard
    $p$-form hierarchy.  This extension involves additional 3- and
    4-forms. One additional 3-form is dual to the superpotential (seen as a
    deformation of the simplest theory).  

    We study the closure of the supersymmetry algebra on all the bosonic
    $p$-form fields of the hierarchy up to duality relations. In order to
    close the supersymmetry algebra without the use of duality relations one
    must construct the hierarchy in terms of supermultiplets. Such a
    construction requires fermionic duality relations among the hierarchy's
    fermions and these turn out to be local.
 }

\end{quotation}

\end{center}

\newpage
\tableofcontents

\newpage
\pagestyle{plain}


\newpage

\section{Introduction}

The embedding tensor formalism\footnote{For recent reviews see
  Refs.~\cite{Trigiante:2007ki,Weidner:2006rp,Samtleben:2008pe}.}, introduced
in
Refs.~\cite{Cordaro:1998tx,deWit:2002vt,deWit:2003hq,deWit:2005hv,deWit:2005ub}
allows the study of the most general gaugings of field theories and, in
particular, of supergravity theories. So far, it has been applied to
maximally- and half-maximally-extended supergravities in various dimensions
\cite{deWit:2004nw,Samtleben:2005bp,Schon:2006kz,de
  Wit:2007mt,Bergshoeff:2007vb,deWit:2008ta}, but not (or, at least, as we are
going to see, not in detail\footnote{The $N=2,d=4$ case has been partially
  studied in Ref.~\cite{de Vroome:2007zd}.}) to supergravities with less
supersymmetry, in particular $N=1,2$ in $d=4$ and minimal supergravities in
$d=5,6$.

A crucial difference between these two cases is that in the former the global
symmetries of the ungauged theories that act on the fermionic fields (the
group of automorphisms of the supersymmetry algebra or R-symmetry) $H_{\rm
  aut}$ also act on the bosonic fields of the theory, while in the latter they
do not. More precisely, if we denote by $G$ the global symmetry group of the
ungauged theories and by $G_{\rm bos}$ the subgroup of $G$ that acts on the
bosons, in the maximal or half-maximal supergravities, $H_{\rm aut} \subset
G_{\rm bos} = G$. In particular, the scalars parametrize the coset $G/(H_{\rm
  aut}\times H_{\rm matter})$ where $H_{\rm matter}$ is related to the matter
multiplets and it is trivial in maximally-extended supergravities.

The situation in $N=1,2$ supergravities in $d=4$ or in minimal supergravities in
$d=5,6$ is totally different: one can write $G=G_{\rm bos} \times H_{\rm
  aut}$. Further, in theories with low amounts of supersymmetry there may exist
symmetries that act only on the vectors (and spinors) but not on the scalars. This
is particularly clear in $N=1,d=4$ supergravity where one can take only vector
supermultiplets and no chiral supermultiplets. The corresponding symmetry group is
the invariance group of the 
    complex vector kinetic matrix. These facts have to be taken into
account properly and it is our goal to do so for the general case of $N=1,d=4$
supergravity by extending the recently found
general 4-dimensional tensor hierarchy \cite{Bergshoeff:2009ph}. The 4-dimensional tensor hierarchy has also been studied in \cite{deWit:2009zv}.

The tensor hierarchy
\cite{deWit:2005hv,deWit:2005ub,deWit:2008ta,Bergshoeff:2009ph,deWit:2009zv}. is an
interesting structure that arises as part of the embedding tensor formalism. It
consists of a system of $p$-form potential fields of all degrees $p=1,\cdots,
d$ in terms of which one can construct gauge-covariant field strengths of all
degrees $p=2,\cdots, d$.  The starting point in the construction of the tensor
hierarchy associated to the gauging of some theory is the field content and
global symmetry group of that theory. This global symmetry group is gauged
using the embedding tensor formalism and in order to have gauge-covariant
field strengths it is usually necessary to introduce higher-rank $p$-form
potentials in a bootstrap procedure ending with the introduction of $p=d$-form
potentials.

The extra $p$-form fields that one has to introduce in the construction of the
hierarchy turn out to be dual to objects such as Noether currents, deformation
parameters etc.~of the field theory and, therefore, do not add new degrees of
freedom when we add them to the theory. Only some of them are completely
necessary to construct a gauge-invariant action for the gauged field
theory. In the $d=4$ case these are the 1- and 2-forms. Then, why should we be
interested in the rest, apart from their need for consistency of the full
construction?

Perhaps the main reason why one should be interested in all the higher-rank
$p$-forms of a theory is the relation between supergravity $p$-form potentials
and supersymmetric $(p-1)$-extended objects (``branes'') which is at the core
of many of the advances made over the last decade in String Theory. While the
branes associated to higher--rank $p$-forms of the 10-dimensional
supergravities are by now well known
\cite{Bergshoeff:2001pv,Bergshoeff:2005ac,Bergshoeff:2006qw,Bergshoeff:2006gs},
little or nothing is known about those of supergravities with lower
supersymmetry and dimensionality like $N=1,2$, $d=4$ supergravity. For
instance, in Ref.~\cite{Bergshoeff:2007ij} it was found that one can
introduce, consistently with supersymmetry, 2-forms in $N=2,d=4$ supergravity
associated with isometries of the scalar manifold to which strings
couple. These 2-forms are ``predicted'' by the 4-dimensional tensor hierarchy
\cite{deWit:2005ub}. But the 4-dimensional tensor hierarchy also predicts
3-forms and 4-forms and one would like to know if they can also be
consistently introduced in the supergravity theory and the kind of extended
objects (domain walls and spacetime-filling branes) they may couple to.

There is another reason to be interested in the higher-rank $p$-forms, in
particular for $p=d-1$ of a supergravity theory. These $(d-1)$-form potentials
are dual to the (``deformation'') parameters that one can introduce
consistently in the theory: gauge coupling constants (represented by the
embedding tensor), St\"uckelberg masses etc. Finding all the $(d-1)$-form
potentials one can get information about the most general deformations
(gaugings, massive deformations...) of the theory.

In this paper we are going to study the possible $p$-form potentials that one
can consistently add to $N=1,d=4$ supergravity, generalizing the results of
Ref.~\cite{Ortin:2008wj}. As we have explained, this problem is related to the
construction of the tensor hierarchy associated to the most general (electric
and magnetic) gauging of the theory. The global symmetry group $G$ of these
theories can be written as\footnote{This splitting of $G$ is factors is not
  unambiguous. In particular, $U(1)_{R}$ transformations can be combined with
  transformations of the other two factors. We will discuss this in detail
  later.} $G_{\rm iso}\times G_{\rm V}\times U(1)_{R}$ where $G_{\rm iso}$ is
the isometry group of the scalar manifold, $G_{\rm V}$ is the invariance group
of the complex vector kinetic matrix and $U(1)_{R}$ the R-symmetry group.  We
will use the general results of \cite{Bergshoeff:2009ph} but supersymmetry
will force us to consider additional fields not contained in the standard
tensor hierarchy. In particular, we will find a 3-form that can be interpreted
as the dual to the superpotential which is a deformation of $N=1,d=4$
supergravity that is not related to a gauging. For earlier work on related
questions see
Refs.~\cite{Bergshoeff:2005ac,Bergshoeff:2006qw,Gomis:2007gb,Bergshoeff:2007ij,Ortin:2008wj}.

This paper is organized as follows: in Section~\ref{sec-N1d4sugra} we review
the standard electric gauging of perturbative symmetries of matter-coupled
$N=1,d=4$ supergravity using the (electric part of the) embedding tensor. This
will allow us to introduce our notation and conventions.  In
Section~\ref{sec-N1d4sugramagnetic} we introduce $N=1,d=4$ supergravity with
electric and magnetic gaugings of perturbative and non-perturbative symmetries
of the theory. This requires the use of the full embedding tensor and the
introduction in the action of the 2-forms predicted by the general
4-dimensional tensor hierarchy. At this point we have a completely consistent
theory with 1- and 2-forms and we do not need to introduce any higher-rank
form potentials unless we worry about the gauge-covariant field strength of
the 2-forms. This is necessary, though, to close (on-shell) the supersymmetry
algebra on the 2-forms and we are led to consider all the $p$-form potentials
predicted by the 4-dimensional tensor hierarchy. We, then, proceed to
construct consistent (on-shell) supersymmetry transformations for all the
hierarchy $p$-form potentials in Section~\ref{sec:susy} which will lead us to
extend the field content of the hierarchy. Finally, we review our results and
present our conclusions in Section~\ref{sec-conclusions}. The appendices
contain summaries of useful formulae concerning K\"ahler geometry and the
4-dimensional tensor hierarchy.


\section{Electrically gauged $N=1, d=4$ supergravity}
\label{sec-N1d4sugra}

In this section we are going to describe the ``standard'' gauged $N=1,d=4$
theory \cite{Cremmer:1982en} using the embedding-tensor formalism. By
``standard'' we mean that only perturbative global symmetries of the ungauged
theory have been gauged using as gauge fields the electric vector fields. In
order to make as clear as possible the construction of the gauged theory, we
are going to describe first the ungauged theory and its global symmetries and
then the gauging procedure.


\subsection{Ungauged $N=1, d=4$ supergravity}
\label{sec-ungaugedN1d4sugra}

The basic\footnote{In the ungauged classical theory (this work is only
  concerned with the classical theory) linear multiplets can always be
  dualized into chiral multiplets and so we do not need to deal with
  them. After the gauging, this is not possible in general, but the embedding
  tensor formalism will allow us to introduce the 2-forms in at a later stage
  in a consistent form.} field content of any $N=1,d=4$ ungauged supergravity
theory is a supergravity multiplet with one graviton $e^{a}{}_{\mu}$ and one
chiral gravitino\footnote{The conventions used here are essentially those of
  Refs.~\cite{Ortin:2008wj} and \cite{kn:toappear2}.}  $\psi_{\mu}$, $n_{C}$
chiral multiplets with as many chiralinos $\chi^{i}$ and complex scalars
$Z^{i}$, $i=1,\cdots, n_{C}$ that parametrize an arbitrary K\"ahler-Hodge
manifold with metric $\mathcal{G}_{ij^{*}}$, and $n_{V}$ vector multiplets
with as many Abelian vector fields $A^{\Lambda}$ with field strengths
$F^{\Lambda}= dA^{\Lambda}$ and chiral gauginos $\lambda^{\Lambda}$,
$\Lambda=1,\cdots,n_{V}$.

In the ungauged theory the couplings between the above fields are determined
by the K\"ahler metric\footnote{The elements of K\"ahler geometry needed in
  this paper are reviewed in Appendix~\ref{app-kahlergeometry}.}
$\mathcal{G}_{ij^{*}}$, an arbitrary holomorphic kinetic matrix
$f_{\Lambda\Sigma}(Z)$ with positive-definite imaginary part and an arbitrary
holomorphic superpotential $W(Z)$ which appears through the covariantly
holomorphic section of K\"ahler weight $(1,-1)$ $\mathcal{L}(Z,Z^{*})$:

\begin{equation}
\mathcal{L}(Z,Z^{*}) = W(Z)e^{\mathcal{K}/2}\, ,
\end{equation}

\noindent
so its K\"ahler-covariant derivative given in
Eq.~(\ref{eq:Kcovariantderivative}) for $\bar{q}=-1$ is
$\mathcal{D}_{i^{*}}\mathcal{L} = e^{\mathcal{K}/2}\partial_{i^{*}}W = 0$.  In
absence of scalar fields, it is possible to introduce a constant
superpotential $\mathcal{L}=W= w$.

The chirality of the spinors is related to their K\"ahler weight: $\psi_{\mu},
\lambda^{\Sigma}$ and $\chi^{i}$ have the same chirality and $\psi_{\mu},
\lambda^{\Sigma}$ and $\chi^{*i^{*}}$ have the same K\"ahler weight
$(1/2,-1/2)$ so their covariant derivatives take the form of
Eq.~(\ref{eq:Kcovariantderivative2}) with $q=1/2$.

The action for the bosonic fields in the ungauged theory is 

\begin{equation}
\label{eq:actionN1u}
 S_{\rm u}  =  {\displaystyle\int} 
\left[
\star R -2\mathcal{G}_{ij^{*}} dZ^{i} 
\wedge\star dZ^{*\, j^{*}}
-2\Im{\rm m}f_{\Lambda\Sigma} 
F^{\Lambda}\wedge \star F^{\Sigma}
+2\Re{\rm e}f_{\Lambda\Sigma} 
F^{\Lambda} \wedge F^{\Sigma}
-\star V_{\rm u} 
\right]\, ,
\end{equation}

\noindent
where the scalar potential $V_{\rm u}$ is given by 

\begin{equation}
\label{eq:Vu}
V_{\rm u}(Z,Z^{*}) = 
-24 |\mathcal{L}|^{2} 
+8\mathcal{G}^{ij^{*}}\mathcal{D}_{i}\mathcal{L} \mathcal{D}_{j^{*}}\mathcal{L}^{*}\, . 
\end{equation}

In absence of scalar fields the constant superpotential $\mathcal{L}=W= w$
leads to an anti-de Sitter-type cosmological constant 

\begin{equation}
\label{eq:Vuw}
V_{\rm u} = -24 |w|^{2}\, . 
\end{equation}

The supersymmetry transformation rules for the fermions (to first order in
fermions) are

\begin{eqnarray}
\delta_{\epsilon}\psi_{\mu} & = & 
\mathcal{D}_{\mu}\epsilon +i\mathcal{L}\gamma_{\mu}\epsilon^{*}
=
\left[\nabla_{\mu} 
+{\textstyle\frac{i}{2}}\mathcal{Q}_{\mu}\right]\epsilon 
+i\mathcal{L}\gamma_{\mu}\epsilon^{*}\, ,
\label{eq:gravisusyruleN1u}\\
& & \nonumber \\
\delta_{\epsilon}\lambda^{\Lambda} & = &
{\textstyle\frac{1}{2}}\!\not\! F^{\Lambda+}\epsilon\, ,
\label{eq:gaugsusyruleN1u} \\
& & \nonumber \\
\delta_{\epsilon}\chi^{i} & = & 
i\!\not\!\partial Z^{i}\epsilon^{*}
+2\mathcal{G}^{ij^{*}}\mathcal{D}_{j^{*}}\mathcal{L}^{*}\epsilon\, .
\label{eq:dilasusyruleN1u}
\end{eqnarray}

\noindent
The last terms in Eqs.~(\ref{eq:gravisusyruleN1u}) and
(\ref{eq:dilasusyruleN1u}) are \textit{fermion shifts} associated to the
superpotential which contribute quadratically to the potential $V_{\rm u}$.

In absence of scalar fields and with constant superpotential $\mathcal{L}=W=
w$ the fermion shift in Eq.~(\ref{eq:gravisusyruleN1u}) can be interpreted as
part of an anti-de Sitter covariant derivative

\begin{equation}
\label{eq:gravisusyruleN1uw}  
\delta_{\epsilon}\psi_{\mu} 
=
\nabla_{\mu}\epsilon +iw\gamma_{\mu}\epsilon^{*}\, .
\end{equation}

The supersymmetry transformation rules for the bosonic fields (to the same
order in fermions) are

\begin{eqnarray}
\label{eq:susytranseN1}
\delta_{\epsilon} e^{a}{}_{\mu} & = &  
-{\textstyle\frac{i}{4}} \bar{\psi}_{\mu}\gamma^{a}\epsilon^{*}
+\mathrm{c.c.}\, ,\\
& & \nonumber \\
\label{eq:susytransAN1}
\delta_{\epsilon} A^{\Lambda}{}_{\mu} & = & 
{\textstyle\frac{i}{8}}
\bar{\lambda}^{\Lambda}\gamma_{\mu}
\epsilon^{*}
+\mathrm{c.c.}\, ,\\
& & \nonumber \\
\delta_{\epsilon} Z^{i} & = & 
{\textstyle\frac{1}{4}} \bar{\chi}^{i}\epsilon\, .
\label{eq:susytransZN1}
\end{eqnarray}


\subsection{Perturbative symmetries of the ungauged theory}
\label{sec-pertursymm}

The possible matter couplings of $N=1,d=4$ supergravities are quite
unrestricted. As a result, the global symmetries of these theories can be very
different from case to case. Depending on the couplings it is possible to
have, at the same time, symmetry transformations that only act on certain
fields and not on the rest and symmetry transformations that act
simultaneously on all of them. Thus, it is not easy to describe all the
possible global symmetry groups in a form that is at the same time unified and
detailed without introducing a very complicated notation with several
different kinds of indices. We are going to try to find an equilibrium between
simplicity and usefulness.

Therefore, we are going to denote the group of all the global symmetries of
the theory we work with\footnote{In this section we will use this notation
  only for the perturbative symmetries and later on we will use the same
  notation for all symmetries. It should be easy to recognize from the context
  which case we are talking about.} by $G$ and its generators by $T_{A}$ with
$A,B,C=1,\cdots, \mathrm{rank}\, G$. They satisfy the Lie algebra 

\begin{equation}
\label{eq:SLiealgebra}
[T_{A},T_{B}]= -f_{AB}{}^{C} T_{C}\, .
\end{equation}

\noindent
We denote by $G_{\rm bos}$ the subgroup of transformations of $G$ that act on
the bosonic fields and its generators by $T_{\rm a}$ with
$\mathrm{a},\mathrm{b},\mathrm{c} = 1,\cdots, \mathrm{rank}\, G_{\rm bos} \leq
\mathrm{rank}\, G$. They satisfy the Lie subalgebra

\begin{equation}
[T_{\rm a},T_{\rm b}]= -f_{\rm ab}{}^{\rm c} T_{\rm c}\, .
\end{equation}

\noindent
In $N=1,d=4$ supergravity we have $G=G_{\rm bos}\times U(1)_{R}$ and
$\mathrm{rank}\, G_{\rm bos} = \mathrm{rank}\, G\, -1$. We split the indices
accordingly as $A=(\mathrm{a},\sharp)$. We may introduce a further splitting
of the indices of $G_{\rm bos}$, $\mathrm{a}=(\mathbf{a},\underline{\rm a})$
to distinguish between those that act on the scalars (holomorphic isometries,
belonging to the group\footnote{Not all the isometries of the metric will be
  perturbative or even non-perturbative symmetries of the full theory. They
  have to satisfy further conditions that we are going to study next. It is
  understood that, in order not to have a complicated notation, we denote by
  $G_{\rm iso}$ only those isometries which really are symmetries of the full
  theory and not the full group of isometries of $\mathcal{G}_{ij^{*}}$
  (although they may eventually coincide).}  $G_{\rm iso}\subset G_{\rm bos}$)
and those that do not. The latter, as we will see, constitute the subgroup
$G_{\rm V}\subset G_{\rm bos}$ of symmetries that only act on the vector
(super)fields and leave invariant the kinetic matrix $f_{\Lambda\Sigma}$. We
have, then, $G_{\rm bos}=G_{\rm iso}\times G_{\rm V}$, since any bosonic
symmetry transformation is either an element of $G_{\rm iso}$ or of $G_{\rm
  V}$ and further since by construction no element of $G_{\rm iso}$ can also
be an element of $G_{\rm V}$ and vice versa.

Let us describe the $U(1)_{R}$ transformations first. Under a $U(1)_{R}$
transformation with constant parameter $\alpha^{\sharp}$, objects with K\"ahler
weight $q$ are multiplied by the phase $e^{-iq\alpha^{\sharp}}$.  All the fermions
$\psi_{\mu},\lambda^{\Sigma},\chi^{*\, i^{*}}$, have a non-vanishing K\"ahler
weight $1/2$, though. All the bosons have zero K\"ahler weight and do not
transform under $U(1)_{R}$. 

The superpotential $\mathcal{L}$ has a non-vanishing K\"ahler weight and
therefore transforms under $U(1)_{R}$. As a general rule, in the presence of a
non-vanishing superpotential, $U(1)_{R}$ will only be a symmetry of $N=1,d=4$
supergravity if the phase factor acquired by $\mathcal{L}$ in a $U(1)_{R}$
transformation can be identified with a $U(1)$ transformation of the scalars
that leaves invariant the rest of the action. These transformations, which are
necessarily isometries of the K\"ahler metric will be described next, but we
can already give two examples to clarify the above statement.

\begin{enumerate}
\item Let us consider the case with no chiral superfields and, therefore, no
  scalars and a constant $\mathcal{L}=W= w$ giving rise to the potential
  Eq.~(\ref{eq:Vuw}) and the gravitino supersymmetry transformation
  Eq.~(\ref{eq:gravisusyruleN1uw}). In this case $U(1)_{R}$ transforms the
  complex constant $w$ into $e^{-i\alpha^{\sharp}}w$ and, therefore it is not a
  symmetry since symmetry transformations act on fields, not on coupling
  constants. Certainly, we can never gauge these transformations since the
  local phases would transform a constant into a function which is not a
  field.

\item Let us consider a theory with just one chiral supermultiplet, with
  K\"ahler potential $\mathcal{K}=|Z|^{2}$ and superpotential $W(Z)=wZ$ where
  $w$ is some complex constant so $\mathcal{L}=wZ e^{|Z|^{2}/2}$. In this case
  $U(1)_{R}$ transforms $\mathcal{L}(Z,Z^{*})$ into
  $\mathcal{L}^{\prime}(Z,Z^{*})=w e^{-i\alpha^{\sharp}} Z e^{|Z|^{2}/2}$. This
  transformation can be seen as a transformation of the scalar $Z^{\prime}=
  e^{-i\alpha^{\sharp}}Z$ which happens to leave invariant the K\"ahler potential,
  metric etc. In this case $U(1)_{R}$ is a symmetry when identified with a $U(1)$
  transformation acting on the complex scalar.

\end{enumerate}

The $G_{\rm iso}$ transformations with constant parameters $\alpha^{\bf a}$
act on the complex scalars $Z^{i}$ as reparametrizations

\begin{equation}
\label{eq:deltazio}
\delta_{\alpha} Z^{i} = \alpha^{\bf a} k_{\bf a}{}^{i}(Z)\, .  
\end{equation}

\noindent
If these transformations are symmetries of the full theory they must, first,
preserve the metric $\mathcal{G}_{ij^{*}}$ and its Hermitean structure, which
implies that the $k_{\bf a}{}^{i}$s are the holomorphic components of a set of
Killing vectors $\{K_{\bf a}= k_{\bf a}{}^{i}\partial_{i} +k^{*}_{\bf
  a}{}^{i^{*}}\partial_{i^{*}}\}$ that satisfy the Lie algebra of the group
$G_{\rm iso}$

\begin{equation}
\label{eq:Liealgebra1}
[K_{\bf a},K_{\bf b}]= -f_{\bf a b}{}^{\bf c} K_{\bf c}\, .
\end{equation}

\noindent
The holomorphic and antiholomorphic components satisfy, separately, the same
Lie algebra. 

We can formally add to this algebra, vanishing ``Killing vectors''
$K_{\underline{\rm a}}$ associated to the transformations that do not act on
the scalars (but do act on the vectors), so we have the full algebra of
$G_{\rm bos}$

\begin{equation}
\label{eq:Liealgebra2}
[K_{\rm a},K_{\rm b}]= -f_{\rm a b}{}^{\rm c} K_{\rm c}\, .
\end{equation}

\noindent
Further, we can also add another vanishing Killing vector $K_{\sharp}$, formally
associated to $U(1)_{R}$ and write the full Lie algebra of $G$

\begin{equation}
\label{eq:Liealgebra}
[K_{A},K_{B}]= -f_{AB}{}^{C} K_{C}\, ,
\end{equation}

\noindent
so the reparametrizations of the scalars $Z^{i}$ can be written as

\begin{equation}
\label{eq:deltazio2}
\delta_{\alpha} Z^{i} = \alpha^{A} k_{A}{}^{i}(Z)\,.
\end{equation}

The Killing property of the reparametrizations only ensures the invariance of
the kinetic term for the scalars. In order to be symmetries of the full theory
they must preserve the entire K\"ahler-Hodge structure and leave invariant the
superpotential and the kinetic terms for the vector fields.

\begin{enumerate}

\item Let us start with the K\"ahler structure. The reparametrizations must
  leave the K\"ahler potential invariant up to K\"ahler transformations, i.e.,
  for each Killing vector $K_{A}$

\begin{equation}
\label{eq:Kconservation}
\pounds_{A}\mathcal{K}\equiv \pounds_{K_{A}}\mathcal{K}=
 k_{A}{}^{i}\partial_{i}\mathcal{K}  
 +
 k^{*}_{A}{}^{i^{*}}\partial_{i^{*}}\mathcal{K} 
 =
 \lambda_{A}(Z)+ \lambda^{*}_{A}(Z^{*})\, .
\end{equation}

\noindent
This relation is consistent for $A=\underline{\rm a},\sharp$, if 

\begin{equation}
\Re{\rm e}\, \lambda_{\underline{\rm a}}=\Re{\rm e}\, \lambda_{\sharp}=0\, .
\end{equation}

\noindent
Furthermore, the reparametrizations must preserve the K\"ahler 2-form
$\mathcal{J}$

\begin{equation}
\label{eq:Jconservation}
\pounds_{A}\mathcal{J}=0\, .
\end{equation}

\noindent
The closedness of $\mathcal{J}$ implies that $\pounds_{A}\mathcal{J} =
d(i_{K_{A}}\mathcal{J})$ and therefore the preservation of the K\"ahler
structure implies the existence of a set of real functions $\mathcal{P}_{A}$
called \textit{momentum maps} such that

\begin{equation}
\label{eq:KMomMap}
  i_{K_{A}}\mathcal{J}= d\mathcal{P}_{A}\, ,
\end{equation}

\noindent
which is also consistent for $A=\underline{\rm a},\sharp$ if the corresponding

\begin{equation}
\label{eq:constantmomaps}
\mathcal{P}_{\underline{\rm a}}=  \mathcal{P}_{\sharp} =\,\, \mathrm{constant}\, .  
\end{equation}

\noindent
Using only Eq.~(\ref{eq:Kconservation}) a local solution to Eq.~(\ref{eq:KMomMap}) is
provided by

\begin{equation}\label{eq:momentummap}
  i\mathcal{P}_{A}
  =
  k_{A}{}^{i}\partial_{i}\mathcal{K} -\lambda_{A}\, ,   
\end{equation}

\noindent
which, on account of Eq.~(\ref{eq:Kconservation}) is equivalent to

\begin{equation}
  i\mathcal{P}_{A}
  =
  -(k^{*}_{A}{}^{i^{*}}\partial_{i^{*}}\mathcal{K} 
  -\lambda^{*}_{A})\, ,
\end{equation}

\noindent
so that, for $A=\underline{\rm a},\sharp$,

\begin{equation}
  \lambda_{\underline{\rm a}}= -i\mathcal{P}_{\underline{\rm a}}\, ,
  \hspace{1cm} 
  \lambda_{\sharp}= -i\mathcal{P}_{\sharp}\, ,
\end{equation}

\noindent
where $\mathcal{P}_{\underline{\rm a}}$ and $\mathcal{P}_{\sharp}$ are real
constants (see Eq.~(\ref{eq:constantmomaps})). Eq.~(\ref{eq:momentummap}) implies that the momentum maps can be used as  prepotentials from which the Killing vectors
can be derived:

\begin{equation}
  \label{eq:prepo}
  k_{A\, i^{*}} =i\partial_{i^{*}}\mathcal{P}_{A}\, .  
\end{equation}

\noindent
Observe that this equation is consistent with the triviality of the ``Killing
vectors'' $K_{\underline{\rm a}},K_{\sharp}$ and the constancy of the
corresponding momentum maps Eq.~(\ref{eq:constantmomaps}). 

Using Eqs.~(\ref{eq:Liealgebra}), (\ref{eq:Kconservation}) and (\ref{eq:momentummap}) it can be shown that the momentum maps satisfy the so-called equivariance condition:

\begin{equation}
\label{eq:momentummaptransformationrule}
\pounds_{A}\mathcal{P}_{B} = 
2i k_{[A}{}^{i}k^{*}_{B]}{}^{j^{*}}\mathcal{G}_{ij^{*}} =
-f_{AB}{}^{C} \mathcal{P}_{C}\, .  
\end{equation}

\noindent
This equivariance condition implies that momentum maps can only be constant and different from zero for Abelian factors. These constants will be associated after gauging to the
\textit{D}- or \textit{Fayet-Iliopoulos} terms.

\item If the K\"ahler-Hodge structure is preserved, any section $\Phi$ of
  K\"ahler weight $(p,q)$ must transform as\footnote{We do not write
    explicitly any spacetime, target space etc.~indices.}

\begin{equation}
\delta_{\alpha}\Phi = -\alpha^{A}(\mathbb{L}_{A}
-K_{A}) \Phi\, ,  
\end{equation}

\noindent
where $\mathbb{L}_{A}$ stands for the symplectic and K\"ahler-covariant Lie
derivative w.r.t.~$K_{A}$ and is given by

\begin{equation}
\mathbb{L}_{A} \Phi \equiv \{\pounds_{A}
+[T_{A}
+{\textstyle\frac{1}{2}}(p\lambda_{A}+q\lambda^{*}_{A})]\}
\Phi\, ,  
\end{equation}

\noindent
where the $T_{A}$ are the matrices that generate the subgroup of $G_{\rm bos}$ that acts on the vectors. The $T_A$ are assumed to be in the representation in which the section transforms and they satisfy the Lie algebra
Eq.~(\ref{eq:SLiealgebra}).  This means that the gravitino $\psi_{\mu}$
transforms according to

\begin{equation}
\delta_{\alpha}\psi_{\mu} = -{\textstyle\frac{i}{2}}\alpha^{A} \Im{\rm m}\, 
\lambda_{A} \psi_{\mu}\, .  
\end{equation}

For $A=\underline{\rm a},\sharp$ we have just $U(1)_{R}$ transformations for each
component $\mathcal{P}_{\underline{\rm a}},\mathcal{P}_{\sharp}$ different from
zero. For $A=\mathbf{a}$ the transformations are still global but the $
\Im{\rm m}\, \lambda_{A}$s are in general functions of $Z,Z^{*}$. These cannot
be compensated by $U(1)_{R}$ transformations.

The chiralinos $\chi^{i}$ transform according to 

\begin{equation}
\delta_{\alpha}\chi^{i} = \alpha^{A} \{ \partial_{j}k_{A}{}^{i}\chi^{j}
+{\textstyle\frac{i}{2}}\Im{\rm m}\, \lambda_{A} \chi^{i}\}\, ,
\end{equation}

\noindent
and the transformations of the gauginos will be discussed after we discuss the
transformations of the vector fields.

\item Let us now consider the invariance of the superpotential $W$. We can
  require, equivalently, that the section $\mathcal{L}$ be invariant up to
  K\"ahler transformations. A K\"ahler-weight $(p,q)$ section $\Phi$ will be
  invariant up to K\"ahler transformations if\footnote{This condition only makes sense for transformations
    $K_{\bf a}$ that really act on the scalars.}

\begin{equation}
\mathbb{L}_{\bf a} \Phi =0\, ,\,\,\,\, \Rightarrow\,\,\,\,  
\pounds_{\bf a}\Phi = 
-[T_{\bf a}
+{\textstyle\frac{1}{2}}(p\lambda_{\bf a}+q\lambda^{*}_{\bf a})]
\Phi\, .
\end{equation}

\noindent
Therefore, we must require for all $A=\mathbf{a}$

\begin{equation}
 K_{\bf a}\mathcal{L} =
-i\Im{\rm m}\, \lambda_{\bf a} \mathcal{L}\, ,
\,\,\,\, \Rightarrow\,\,\,\,
\delta_{\alpha}\mathcal{L} = -i\alpha^{\bf a}\Im{\rm m}\, \lambda_{\bf a} \mathcal{L}\, ,
\end{equation}

\noindent
but we cannot extend straightforwardly the same expression to all $A$ since,
as discussed at the beginning of this section, the corresponding
transformations (constant phase multiplications) are only symmetries when
$\mathcal{L}=0$ or when they are associated to transformations of the scalars
and this is, by definition, not the case when $A=\underline{\rm a},\sharp$. 

We, therefore, write 

\begin{equation}
\delta_{\alpha}\mathcal{L} = -i\alpha^{A}\Im{\rm m}\, \lambda_{A} \mathcal{L}\, ,
\end{equation}

\noindent
imposing at the same time the constraint\footnote{This constraint should be
  understood as a way to consider the cases $\mathcal{L}=0$ and
  $\mathcal{L}\neq 0$ simultaneously: when $\mathcal{L}\neq 0$ the symmetry
  transformations must satisfy $(\alpha^{\underline{\rm
      a}}\mathcal{P}_{\underline{\rm a}} +
  \alpha^{\sharp}\mathcal{P}_{\sharp})=0$ and they are unrestricted when
  $\mathcal{L}=0$.}

\begin{equation}
\label{eq:superpotentialconstraint1}  
(\alpha^{\underline{\rm a}}\Im{\rm m}\, \lambda_{\underline{\rm a}} 
+  
\alpha^{\sharp}\Im{\rm m}\, \lambda_{\sharp})\mathcal{L}=
(\alpha^{\underline{\rm a}}\mathcal{P}_{\underline{\rm a}} 
+  
\alpha^{\sharp}\mathcal{P}_{\sharp})\mathcal{L}
=
0\, .
\end{equation}

\item The kinetic term for the vector fields $A^{\Lambda}$ in the action will
  be invariant\footnote{It is at this point that the restriction to
    perturbative symmetries (symmetries of the action) is made.} if the effect
  of a reparametrization on the kinetic matrix $f_{\Lambda\Sigma}$ is
  equivalent to a rotation on its indices that can be compensated by a
  rotation of the vectors, or a constant Peccei-Quinn-type shift i.e.

\begin{eqnarray}
\delta_{\alpha} f_{\Lambda\Sigma} & \equiv  & -\alpha^{\rm a}\pounds_{\rm a}
f_{\Lambda\Sigma}
= 
\alpha^{\rm a}[T_{\mathrm{a}\, \Lambda\Sigma}
-2  T_{\mathrm{a}\, (\Lambda}{}^{\Omega}f_{\Sigma)\Omega}]\, ,\\
& & \nonumber \\
\delta_{\alpha}A^{\Lambda} & = & \alpha^{\rm a} T_{\mathrm{a}\,
  \Sigma}{}^{\Lambda}A^{\Sigma}\, ,
\end{eqnarray}

\noindent
where the shift generator is symmetric $T_{\mathrm{a}\, \Lambda\Sigma}=
T_{\mathrm{a}\, \Sigma\Lambda}$ to preserve the symmetry of the kinetic
matrix. 

Observe that for $\mathrm{a}=\underline{\rm a}$,
    $\pounds_{\underline{\rm a}}f_{\Lambda\Sigma}=0$, and, for consistency, we
    must have $T_{\underline{\rm a}\,
      (\Lambda}{}^{\Omega}f_{\Sigma)\Omega}=0$, i.e.~the transformations
    $T_{\underline{\rm a}}$ are those that preserve the kinetic matrix. This is why we call the 
    group generated by $T_{\underline{\rm a}}$ the invariance group $G_{\rm V}$ of the 
    complex vector kinetic matrix.

The iteration of two of these infinitesimal transformations indicates that
they can be described by the $2n_{V}\times 2n_{V}$ matrices\footnote{Observe
  that this group is the semidirect product of the group that rotates the
  vectors, generated by the matrices $T_{\mathrm{a}\, \Sigma}{}^{\Lambda}$ and
  the Abelian group of shifts generated by the matrices $T_{\mathrm{a}\,
    \Lambda\Sigma}$. Evidently, some of these matrices identically
  vanish. This is the price we have to pay to use the same indices
  $\mathrm{a},\mathrm{b},\mathrm{c},\ldots$ for the generators of both
  groups.}

\begin{equation}
\label{eq:triangulartransf}
T_{\mathrm{a}} \equiv 
\left(
  \begin{array}{cc}
 T_{\mathrm{a}\, \Lambda}{}^{\Sigma} & 0 \\
& \\
T_{\mathrm{a}\, \Lambda\Sigma} & T_{\mathrm{a}}{}^{\Lambda}{}_{\Sigma} \\
  \end{array}
\right)\, ,  
\hspace{1cm}
T_{\mathrm{a}}{}^{\Lambda}{}_{\Sigma} \equiv - T_{\mathrm{a}\, \Sigma}{}^{\Lambda}\, ,
\end{equation}

\noindent
satisfying the Lie algebra

\begin{equation}
[T_{\mathrm{a}}, T_{\mathrm{b}}]= -f_{\mathrm{a}\mathrm{b}}{}^{\mathrm{c}}
T_{\mathrm{c}}\, .  
\end{equation}

\noindent
As we have discussed some of the transformations generated by the $K_{\bf a}$
may only act on the scalars and not on the vectors, for instance, because the
kinetic matrix does not depend on the relevant scalars. We assume that the
corresponding subset of $2n_{V}\times 2n_{V}$ matrices $T_{\bf a}$ are
identically zero. On the other hand, we can formally add to these matrices
another identically vanishing $2n_{V}\times 2n_{V}$ matrix $T_{\sharp}$ so we have
a full set of $2n_{V}\times 2n_{V}$ matrices $T_{A}$ satisfying the Lie
algebra of $G$, Eq.~(\ref{eq:SLiealgebra}).

\end{enumerate}

Combining all these results we conclude that the gauginos transform according
to 

\begin{equation}
\delta_{\alpha}\lambda^{\Sigma} = -\alpha^{A} [T_{A\,
  \Omega}{}^{\Sigma}\lambda^{\Omega} 
+{\textstyle\frac{i}{2}} \Im{\rm m}\, 
\lambda_{A} \lambda^{\Sigma}]
\, .
\end{equation}

At this point there is no restriction on the group $G$ nor on the $n_{V}\times
n_{V}$ matrices $T_{A\, \Lambda}{}^{\Sigma}$, although one can already see
that the lower-triangular $2n_{V}\times 2n_{V}$ matrices $T_{A}$ are
generators of the symplectic group.


\subsection{Electric gaugings of perturbative symmetries}

We are now going to gauge the symmetries described in the previous subsection
using as gauge fields the electric 1-form potentials $A^{\Lambda}$. This
requires the introduction of the (electric) \textit{embedding tensor}
$\vartheta_{\Lambda}{}^{A}$ to indicate which global symmetry is
gauged by which gauge field $A^{\Lambda}$ and, equivalently, to identify the
parameters of global symmetries $\alpha^{A}$ that are going to be promoted to
local parameters with the gauge parameters $\Lambda^{\Sigma}(x)$ of the
1-forms:

\begin{equation}\label{eq:gaugetrafoparameter}
\alpha^{A}(x)\equiv \Lambda^{\Sigma}(x)\vartheta_{\Sigma}{}^{A}\, .  
\end{equation}

We will write now the constraint Eq.~(\ref{eq:superpotentialconstraint1}) in
the form\footnote{Again, this constraint and other constraints of the same
  kind that will follow, should be understood as a way to consider the cases
  $\mathcal{L}=0$ and $\mathcal{L}\neq 0$ simultaneously: when
  $\mathcal{L}\neq 0$ the embedding tensor must satisfy
  $(\vartheta_{\Sigma}{}^{\underline{\rm a}}\mathcal{P}_{\underline{\rm a}} +
  \vartheta_{\Sigma}{}^{\sharp}\mathcal{P}_{\sharp})=0$ and it is unrestricted
  when $\mathcal{L}=0$.}

\begin{equation}
\label{eq:superpotentialconstraint2}  
(\vartheta_{\Sigma}{}^{\underline{\rm a}}\mathcal{P}_{\underline{\rm a}} 
+  
\vartheta_{\Sigma}{}^{\sharp}\mathcal{P}_{\sharp})\mathcal{L}=0\, .
\end{equation}

Taking into account Eq.~(\ref{eq:deltazio2}) and the definition Eq.~(\ref{eq:gaugetrafoparameter}), the gauge
transformations of the complex scalars will be 

\begin{equation}
\label{eq:deltazel}
\delta Z^{i} = \Lambda^{\Sigma}\vartheta_{\Sigma}{}^{A}k_{A}{}^{i}\, .  
\end{equation}

The embedding tensor cannot be completely arbitrary. To start with, it is
clear that it has to be invariant under gauge transformations, which we denote
by $\delta$: 

\begin{equation}
\label{eq:quadraticTdefel}
\delta \vartheta_{\Lambda}{}^{A} =   
- \Lambda^{\Sigma} Q_{\Sigma \Lambda}{}^{A}\, , 
\hspace{1cm}
Q_{\Sigma \Lambda}{}^{A}
\equiv 
\vartheta_{\Sigma}{}^{B}T_{B\,  \Lambda}{}^{\Omega}\vartheta_{\Omega}{}^{A}
-\vartheta_{\Sigma}{}^{B}\vartheta_{\Lambda}{}^{C}f_{BC}{}^{A}\, .  
\end{equation}

\noindent
Then, the embedding tensor has to satisfy the quadratic constraint

\begin{equation}
\label{eq:quadraticTel}
Q_{\Sigma\Lambda}{}^{A}=0\, .
\end{equation}

The gauge fields $A^{\Lambda}$ effectively couple to the generators

\begin{equation}
X_{\Sigma \Omega}{}^{\Gamma} \equiv \vartheta_{\Sigma}{}^{A}
T_{A\,  \Omega}{}^{\Gamma}\, ,
\hspace{1cm}
X_{\Sigma \Omega\Gamma} \equiv \vartheta_{\Sigma}{}^{A}
T_{A\,  \Omega\Gamma}\, ,
\hspace{1cm}
X_{\Sigma} \equiv \vartheta_{\Sigma}{}^{A}T_{A}\, .
\end{equation}

\noindent
From the definition of the quadratic constraint Eq.~(\ref{eq:quadraticTel}) 

\begin{equation}
X_{(\Lambda\Sigma)}{}^{\Omega}\vartheta_{\Omega}{}^{A} = 0\, , 
\end{equation}

\noindent
which, for this purely electric gauging case implies

\begin{equation}
X_{(\Lambda\Sigma)}{}^{\Omega} = 0\, , 
\end{equation}

\noindent
and no need to intoduce 2-form potentials.  From the commutator of the
matrices $T_{A}$ and using the quadratic constraint we find the commutator of
$X$ generators

\begin{equation}
\label{eq:unconstrainedXcommutator}
[X_{\Lambda},X_{\Sigma}] = -X_{\Lambda\Sigma}{}^{\Omega}X_{\Omega}\, ,  
\end{equation}

\noindent
from which we can derive the analogue of the Jacobi identities.

We are now ready to gauge the theory. We will not attempt to give the full
supersymmetric Lagrangian and supersymmetry transformation rules, but only
those elements that allow its construction to lowest order in fermions (that is we consider supersymmetry transformations acting on fermions up to first order fermion terms  and supersymmetry transformations acting on bosons up to second order fermions terms).

First, we have to replace the partial derivatives of the scalars in their
kinetic term by the covariant derivatives

\begin{equation}
\mathfrak{D}Z^{i}\equiv dZ^{i}
+A^{\Lambda}\vartheta_{\Lambda}{}^{A}k_{A}{}^{i}\, ,  
\end{equation}

\noindent
where the gauge potentials transform according to

\begin{equation}
\delta A^{\Sigma} = -\mathfrak{D}\Lambda^{\Sigma} \equiv 
-(d\Lambda^{\Sigma} +X_{\Lambda\Omega}{}^{\Sigma}A^{\Lambda}\Lambda^{\Omega})\, .
\end{equation}

\noindent
We also replace in the action the vector field strengths by the
gauge-covariant field strengths

\begin{equation}\label{eq:perturbativegaugecovfieldstrength}
F^{\Sigma} = dA^{\Sigma} +{\textstyle\frac{1}{2}}
X_{\Lambda\Omega}{}^{\Sigma}A^{\Lambda}\wedge A^{\Omega}\, .
\end{equation}

Observe that we have not introduced a coupling constant $g$ as it is standard
in the literature since the embedding tensor already plays the role of
a coupling constant and even of different coupling constants if we are dealing with
products of groups.  Observe also that $\vartheta_{\sharp}{}^{A}$ does not appear
in any of these expressions because $K_{\sharp}=T_{\sharp}=0$.

We have to replace the (K\"ahler- and Lorentz-) covariant derivatives $\mathcal{D}$ of the
spinors in their kinetic terms by the gauge-covariant derivatives $\mathfrak{D}$:

\begin{eqnarray}
\mathfrak{D}_{\mu}\psi_{\nu} 
& = & 
\{\mathcal{D}_{\mu}-{\textstyle\frac{i}{2}}A^{\Lambda}{}_{\mu}\vartheta_{\Lambda}{}^{A}
\mathcal{P}_{A}\}\psi_{\nu}\, ,\\
& & \nonumber \\
\mathfrak{D}\chi^{i} 
& = & 
\mathcal{D}\chi^{i} +\Gamma_{jk}{}^{i}\mathfrak{D}Z^{j} \chi^{k}
-A^{\Lambda}\vartheta_{\Lambda}{}^{A}\partial_{j}k_{A}{}^{i}\chi^{j}
+{\textstyle\frac{i}{2}}A^{\Lambda}\vartheta_{\Lambda}{}^{A}
\mathcal{P}_{A}\chi^{i} \, ,\\
& & \nonumber \\
\mathfrak{D}\lambda^{\Sigma} 
& = & 
\{\mathcal{D}-{\textstyle\frac{i}{2}}A^{\Lambda}\vartheta_{\Lambda}{}^{A}
\mathcal{P}_{A}\}\lambda^{\Sigma} 
-X_{\Lambda\Omega}{}^{\Sigma}A^{\Lambda}\lambda^{\Omega}\, .
\end{eqnarray}

When $\mathcal{L}=0$ the components $\vartheta_{\Lambda}{}^{\sharp}$ and $\vartheta_{\Lambda}{}^{\underline{\rm a}}$ occur in all these covariant derivatives. When $\mathcal{L}\neq 0$ the embedding tensor $\vartheta_{\Lambda}{}^{\sharp}$ does not appear (and $\vartheta_\Lambda{}^{\underline{a}}$ only appears in the last term of $\mathfrak{D}\lambda^{\Sigma}$). In the case $\mathcal{L}\neq 0$ the gauging of the $U(1)_R$ symmetry requires $U(1)_R$ to be identified with a $U(1)$ subgroup acting on the scalars. Thus the embedding tensor component associated to a $U(1)_R$ gauging is contained in $\vartheta_\Lambda{}^{\mathbf{a}}$.

The supersymmetry transformations of the bosonic fields do not change with the
gauging, but those of the fermions do by the above replacement of (K\"ahler- and Lorentz-) covariant derivatives by  gauge-covariant derivatives. Further in the gaugino supersymmetry transformation the field strength is given by Eq.~(\ref{eq:perturbativegaugecovfieldstrength}) and there appears a new fermion shift
term $\mathcal{D}^\Sigma$. To first order in
fermions, we have

\begin{eqnarray}
\delta_{\epsilon}\psi_{\mu} & = & 
\mathfrak{D}_{\mu}\epsilon+i\mathcal{L}\gamma_{\mu}\epsilon^{*}\, ,
\label{eq:gravisusyruleN1}\\
& & \nonumber \\
\delta_{\epsilon}\lambda^{\Sigma} & = &
{\textstyle\frac{1}{2}}\left[\not\!
  F^{\Sigma +}+i\mathcal{D}^{\Sigma}\right]\epsilon\, ,
\label{eq:gaugsusyruleN1} \\
& & \nonumber \\
\delta_{\epsilon}\chi^{i} & = & 
i\not\!\!\mathfrak{D} Z^{i}\epsilon^{*}
+2\mathcal{G}^{ij^{*}}\mathcal{D}_{j^{*}}\mathcal{L}^{*}\epsilon\, ,
\label{eq:dilasusyruleN1}
\end{eqnarray}

\noindent
where $\not\!\!F^{\Sigma +}=\gamma^\mu\gamma^\nu F^{\Sigma +}{}_{\mu\nu}$ in which $F^{\Sigma +}=\tfrac{1}{2}\left(F^{\Sigma}+i\star F^{\Sigma}\right)$ is the selfdual field strength, and 

\begin{equation}
\label{eq:fermionshiftsn1}
\mathcal{D}^{\Lambda} \equiv 
-\Im {\rm m}\, f^{\Lambda\Sigma}\vartheta_{\Sigma}{}^{A}\mathcal{P}_{A}\, ,
\end{equation}

\noindent
where we use the notation

\begin{equation}
\Im {\rm m}\, f^{\Lambda\Sigma} \equiv   (\Im {\rm m}\,
f)^{-1|\Lambda\Sigma}\, .
\end{equation}

\noindent
The new term $\mathcal{D}^\Lambda$ leads to corrections of the scalar potential of the ungauged
theory $V_{\rm u}$, given in Eq.~(\ref{eq:Vu}), which now takes the form

\begin{equation}
  V_{\rm eg} = V_{\rm u}  -\mathcal{D}^{\Lambda}\vartheta_{\Lambda}{}^{A}  \mathcal{P}_{A}
  =V_{\rm u}
  +{\textstyle\frac{1}{2}}
  \Im {\rm m}\,f^{\Lambda\Sigma}\vartheta_{\Lambda}{}^{A}\vartheta_{\Sigma}{}^{B}
  \mathcal{P}_{A}\mathcal{P}_{B}\, . 
\end{equation}

The action for the bosonic fields of the $N=1,d=4$ gauged supergravity of the
kind considered here is obtained by replacing the partial derivatives and
field strengths by gauge-covariant derivatives and field strengths, replacing
the potential $V_{\rm u}$ by $V_{\rm eg}$ above and by adding a Chern--Simons
term \cite{de Wit:1984px,Andrianopoli:2004sv} which is necessary to make the
action gauge invariant

\begin{equation}
\label{eq:actionN1}
\begin{array}{rcl}
 S_{\rm eg}  & =  &
{\displaystyle\int}
\left\{
\star R -2\mathcal{G}_{ij^{*}} \mathfrak{D}Z^{i} 
\wedge\star \mathfrak{D}Z^{*\, j^{*}}
-2\Im{\rm m}f_{\Lambda\Sigma} 
F^{\Lambda}\wedge \star F^{\Sigma}
+2\Re{\rm e}f_{\Lambda\Sigma} 
F^{\Lambda} \wedge F^{\Sigma}
\right.
\\
& & \\
& & 
\left.
-\star V_{\rm eg} 
-{\textstyle\frac{4}{3}}  
X_{\Lambda\Sigma\Omega} A^{\Lambda}\wedge A^{\Sigma} \wedge [dA^{\Omega} 
+{\textstyle\frac{3}{8}} X_{\Gamma\Delta}{}^{\Omega}A^{\Gamma}\wedge A^{\Delta}]
\right\}\, .
\end{array}
\end{equation}

\noindent
Gauge-invariance can be achieved only if 

\begin{equation}
\label{eq:XLSO}
X_{(\Lambda\Sigma\Omega)} =0\, , 
\end{equation}

\noindent
which is a constraint that also follows from supersymmetry.


\section{Electrically and magnetically gauged $N=1, d=4$ supergravity}
\label{sec-N1d4sugramagnetic}

In this section we will discuss the most general gaugings of $N=1, d=4$
supergravity by using as gauge group any subgroup of $G=G_{\rm iso}\times G_{\rm
  V}\times U(1)_{R}$ that can be embedded into $Sp(2n_{\rm
  V},\mathbb{R})$. 

From the purely bosonic point of view it would suffice to use the results of
Refs.~\cite{deWit:2005ub,Bergshoeff:2009ph} taking into account the particular
structure of the global symmetry group of $N=1,d=4$ supergravity. This
involves the introduction of new $p$-form fields $p=2,3,4$ which, together
with the electric and magnetic (to be defined) 1-forms of the theory, combined
into $A^{M}$, constitute the standard 4-dimensional tensor hierarchy, reviewed
in Appendices~\ref{app-projectors} and~\ref{app-gaugetranshierarchy}. Its
field content is

\begin{displaymath}
\{A^{M},B_{A},C_{A}{}^{M},D_{AB},D_{E}{}^{NP}, D^{NPQ}\}\, .
\end{displaymath}

\noindent
At the level of the action, it is not necessary to introduce all these fields,
though. It is enough to introduce the magnetic 1-forms $A_{\Lambda}$ and
2-forms $B_{A}$.

This procedure, however, must be compatible with $N=1,d=4$ supersymmetry. A
supersymmetrization of the tensor hierarchy and the action is necessary. The
supersymmetrization of the tensor hierarchy is a first step towards the
construction of a fully supersymmetric action with electric and magnetic
gaugings and this is going to be our goal in this section.

Thus, we are going to repeat the construction of the 4-dimensional tensor
hierarchy checking at each step its consistency with $N=1,d=4$ supersymmetry:
for each new $p$-form field we will construct a supersymmetry transformation
and we will check the closure of the local $N=1,d=4$ supersymmetry algebra on
it. The commutator of two $N=1,d=4$ local supersymmetry transformations acting
on bosonic $p$-form fields is expected to have the general form

\begin{equation}
\label{eq:susyalgebra2}
\left[\delta_{\eta}\,,\delta_{\epsilon}\right]= 
\delta_{\mathrm{g.c.t.}}+\delta_{\mathrm{gauge}}+\text{duality relations}
\, ,
\end{equation}

\noindent
where $\delta_{\mathrm{g.c.t.}}$ is a general coordinate transformation and
$\delta_{\mathrm{gauge}}$ is a gauge transformation that should coincide with
the one predicted by the bosonic tensor hierarchy purely on the basis of gauge-invariance
arguments. We also expect in general additional terms proportional to duality
relations between the new fields and the original fields of the ungauged
$N=1,d=4$ supergravity. These duality relations project the tensor hierarchy
onto the physical theory reducing the number of independent fields.

Contrary to that expectation, we are going to see that, at least for some
fields, it is possible to construct supersymmetry transformations such that
the local $N=1,d=4$ supersymmetry algebra  closes without the use of any
duality relation, i.e. 

\begin{equation}
\label{eq:susyalgebra}
\left[\delta_{\eta}\,,\delta_{\epsilon}\right]= 
\delta_{\mathrm{g.c.t.}}+\delta_{\mathrm{gauge}}
\, .
\end{equation}

\noindent
To make this possible we will have to introduce the additional $p$-form fields
of the tensor hierarchy in supermultiplets constructing, as a matter of fact,
a supersymmetric tensor hierarchy. Now, to project the supersymmetric tensor
hierarchy onto the physical theory we will use duality relations both for the
bosons and fermions.  

We have succeeded in supersymmetrizing in this way the hierarchy up to 2-forms
(which requires the introduction of linear multiplets) but these results
strongly indicate that the same should be possible for all $p$-forms in the
tensor hierarchy.

Studying the closure of the local $N=1,d=4$ supersymmetry algebra we are going
to see that it is necessary to add more bosonic $p$-form fields to the
standard tensor hierarchy. The main reason for this is the existence of the
constraint Eq.~(\ref{eq:superpotentialconstraint2}) which will be 
generalized to the electric-magnetic case in
Eq.~(\ref{eq:superpotentialconstraint3}).  This constraint restricts
simultaneously the terms $\mathcal{P}_{\underline{\rm
    a}},\mathcal{P}_{\sharp}$ and the symmetries that can be gauged and
reflects the breaking of the $U(1)_{R}$ symmetry by the presence of a
non-vanishing superpotential $\mathcal{L}$.

The breaking of this symmetry will manifest itself in the existence of a new
St\"uckelberg shift of the 2-forms $B_{\underline{\rm a}},B_{\sharp}$

\begin{equation}
\delta B_{\underline{\rm a}} \sim \mathcal{P}_{\underline{\rm a}}\Lambda\, ,
\hspace{1cm}  
\delta B_{\sharp} \sim  \mathcal{P}_{\sharp}\Lambda\, , 
\end{equation}

\noindent
where $\Lambda$ is a 2-form that appears whenever $\mathcal{L}\neq 0$. We can
only find this shift by studying the closure of the local supersymmetry
algebra. Therefore, it is necessary to simultaneously construct the tensor
hierarchy and study its supersymmetrization.

To construct the respective gauge-covariant 3-form field strengths
$H_{\underline{\rm a}}$, $H_{\sharp}$ the existence of one new 3-form $C$ is
required. We will find consistent supersymmetry transformations for the needed
3-form $C$ (as well as for yet another 3-form $C^{\prime}$ that is dual to the superpotential). In order to have 
gauge-covariant 4-form field strengths
$G_{\underline{\rm a}}{}^M$ and $G_{\sharp}{}^M$ we need to introduce a set of 4-forms
$D^{M}$. The extended hierarchy of $N=1,d=4$ supergravity will, thus, have the
total bosonic field content

\begin{displaymath}
\{A^{M}, B_{A}, C_{A}{}^{M}, C, C^{\prime},D_{AB}, D_{E}{}^{NP},
D^{NPQ}, D^{M}\}\, .
\end{displaymath}

We start by reviewing the non-perturbative symmetries of the ungauged theory.


\subsection{Non-perturbative symmetries of the ungauged theory}

The new, non-perturbative symmetries to be considered are symmetries of the
``extended'' equations of motion of the ungauged theory which are the standard
equations of motion plus the Bianchi identities of the vector field strengths:

\begin{equation}
\label{eq:BianchiFL}
dF^{\Lambda}=0\, .  
\end{equation}

\noindent
The Maxwell equations that one obtains from the action
Eq.~(\ref{eq:actionN1u}) can be written as Bianchi identities for the 2-forms
$G_{\Lambda}$

\begin{equation}
\label{eq:MaxwellGl}
dG_{\Lambda}=0\, ,  
\hspace{1.5cm}
G_{\Lambda}{}^{+} \equiv f_{\Lambda\Sigma}(Z)F^{\Sigma\, +}\, .
\end{equation}

This set of extended equations of motion (Maxwell equations plus Bianchi
identities) is invariant under general linear transformations

\begin{equation}
\left(
  \begin{array}{c}
   F^{\Lambda} \\
   G_{\Lambda} \\ 
  \end{array}
\right)^{\prime}  
=
\left(
  \begin{array}{cc}
   A_{\Sigma}{}^{\Lambda} & B^{\Sigma\Lambda} \\
   C_{\Sigma\Lambda} & D^{\Sigma}{}_{\Lambda} \\ 
  \end{array}
\right)
\left(
  \begin{array}{c}
   F^{\Sigma} \\
   G_{\Sigma} \\ 
  \end{array}
\right)\, .
\end{equation}

However, consistency with the definition of $G_{\Lambda}$
Eq.~(\ref{eq:MaxwellGl}) requires that the kinetic matrix transforms at the
same time as

\begin{equation}
\label{eq:fraclinear}
f^{\prime} =(C+Df)(A+Bf)^{-1}\, .  
\end{equation}

\noindent
Then $f^{\prime}$ will be symmetric if 

\begin{equation}
A^{T}C-C^{T}A=0\, ,
\hspace{1cm}  
B^{T}D-D^{T}B=0\, ,
\hspace{1cm}  
A^{T}D-C^{T}B= \xi \mathbb{I}_{n_{V}\times n_{V}}\, ,
\end{equation}

\noindent
where $\xi$ is a constant whose value is found to be $\xi=1$ by the
requirement of invariance of the Einstein equations. 

These conditions can be reexpressed in a better form after introducing some
notation. We define the contravariant tensor of 2-forms $G^{M}$, the
symplectic metric $\Omega_{MN}$ and its inverse $\Omega^{MN}$ which we will
use to, respectively, lower and raise indices

\begin{equation}
G^{M}
\equiv   
\left(
  \begin{array}{c}
F^{\Lambda} \\ G_{\Lambda} \\ 
\end{array}
\right)\, ,
\hspace{1cm}
\Omega_{MN}=\left(
\begin{array}{cc}
0 & \mathbb{I}_{n_{V}\times n_{V}} \\ 
-\mathbb{I}_{n_{V}\times n_{V}} & 0 \\
\end{array}
\right)\, ,  
\hspace{1cm}
\Omega^{MN}\Omega_{NP}=-\delta^{M}{}_{P}\, .
\end{equation}

\noindent
Then, the Maxwell equations and Bianchi identities are formally invariant
under the transformations

\begin{equation}
G^{\prime M}  \equiv M_{N}{}^{M}G^{N}\, ,
\hspace{1cm}
M= (M_{N}{}^{M}) = 
\left(
  \begin{array}{cc}
   A & B \\
   C & D \\ 
  \end{array}
\right)\, ,
\end{equation}

\noindent
satisfying 

\begin{equation}
M^{T}\Omega M = \Omega\, .  
\end{equation}

\noindent
i.e.~$M\in Sp(2n_{V},\mathbb{R})$ \cite{Gaillard:1981rj}.
Infinitesimally\footnote{We include identically vanishing generators
  associated to $U(1)_{R}$ etc. On the other hand, it is clear that the index
  $A$ refers now to more symmetries than in the perturbative case.}

\begin{equation}
M_{N}{}^{M} \sim \mathbb{I}_{2n_{V}\times 2n_{V}} 
+\alpha^{A}T_{A\, N}{}^{M}
= \alpha^{A}
\left(
  \begin{array}{cc}
   T_{A\, \Sigma}{}^{\Lambda} & T_{A}{}^{\Sigma\Lambda} \\
& \\
   T_{A\, \Sigma\Lambda} & T_{A}{}^{\Sigma}{}_{\Lambda} \\ 
  \end{array}
\right)\, ,
\end{equation}

\noindent
and the condition $M\in Sp(2n_{V},\mathbb{R})$ reads

\begin{equation}
\label{eq:Tsymplectic} 
T_{A\, [MN]}\equiv T_{A\, [M}{}^{P}\Omega_{N]P} =0\, . 
\end{equation}

These transformations change the kinetic matrix and will only be symmetries of
all the extended equations of motion if they can be compensated by
reparametrizations, i.e.~$f_{\Lambda\Sigma}$ has to satisfy

\begin{eqnarray}
 \alpha^{A}k_{A}{}^{i}\partial_{i}f_{\Lambda\Sigma}  = 
\alpha^{A}\{ -T_{A\, \Lambda\Sigma}
+2T_{A\, (\Lambda}{}^{\Omega}f_{\Sigma) \Omega}
-T_{A}{}^{\Omega\Gamma}f_{\Omega\Lambda}f_{\Gamma\Sigma}\}\, .
\end{eqnarray}

The subalgebra of matrices that generate symmetries of the action
(perturbative symmetries) are those with $T_{A}{}^{\Sigma\Lambda}=0$, i.e.~the
lower-triangular matrices of Eq.~(\ref{eq:triangulartransf}).

Observe that the transformations acting on the vectors are constrained to
belong to $Sp(2n_{V},\mathbb{R})$. That this is a constraint follows from the fact that the global
symmetry group $G$ is in general not a subgroup of $Sp(2n_{V},\mathbb{R})$. We can thus only gauge those subgroups of $G$ that can be embedded in $Sp(2n_{V},\mathbb{R})$. 

The transformation rule of the kinetic matrix $f_{\Lambda\Sigma}\equiv
R_{\Lambda\Sigma}+iI_{\Lambda\Sigma}$ Eq.~(\ref{eq:fraclinear}) can be
alternatively expressed using the $Sp(2n_{V},\mathbb{R})$ matrix

\begin{equation}
\label{eq:Mdef}
\left( \mathcal{M}^{MN}\right)  
\equiv
\left(
  \begin{array}{cc}
  I^{\Lambda\Sigma}  & I^{\Lambda\Omega}R_{\Omega\Sigma} \\
& \\
R_{\Lambda\Omega}I^{\Omega\Sigma}\hspace{.5cm} & I_{\Lambda\Sigma} 
+R_{\Lambda\Omega}I^{\Omega\Gamma}R_{\Gamma\Sigma} \\
  \end{array}
\right)\, ,
\hspace{1cm}
I^{\Lambda\Omega}I_{\Omega\Sigma}= \delta^{\Lambda}{}_{\Sigma}\, ,
\end{equation}

\noindent
which transforms linearly 

\begin{equation}
\mathcal{M}^{\prime} = M\mathcal{M}M^{T}\, .  
\end{equation}


\subsection{General gaugings of  $N=1, d=4$ supergravity}

We now want to consider the most general gauging of $N=1, d=4$ supergravity,
using perturbative and non-perturbative global symmetries and using electric
and magnetic vectors, to be introduced next. In the ungauged theory we can
introduce $n_{V}$ 1-form potentials $A_{\Lambda}$ and their field strengths
$F_{\Lambda}=dA_{\Lambda}$. The Maxwell equations can be replaced by the
first-order duality relation

\begin{equation}
G_{\Lambda} = F_{\Lambda}\, ,  
\end{equation}

\noindent
since now the Bianchi identity $dF_{\Lambda}=0$ implies the standard Maxwell
equation $dG_{\Lambda}=0$. The magnetic vectors $A_{\Lambda}$ will be
introduced in the theory as auxiliary fields and we will study them from the
supersymmetry point of view later on.  The electric $A^{\Lambda}$ and magnetic
$A_{\Lambda}$ vectors will be combined into a symplectic vector $A^{M}$

\begin{equation}
A^{M}
\equiv   
\left(
  \begin{array}{c}
A^{\Lambda} \\ A_{\Lambda} \\ 
\end{array}
\right)\, ,
\hspace{1cm}
A_{M} \equiv \Omega_{MN}A^{N} = (A_{\Lambda}\, ,
-A^{\Lambda})\, , 
\hspace{1cm}
A^{M} = A_{N}\Omega^{NM}\, ,
\end{equation}

\noindent
and used as the gauge fields of the symmetries described in the previous subsection.

In order to use all the 1-forms $A^{M}$ as gauge fields we need to add a
magnetic component to the embedding tensor, which becomes a covariant
symplectic vector

\begin{equation}
\vartheta_{M}{}^{A}\equiv (\vartheta^{\Lambda A}\, , \vartheta_{\Lambda}{}^{A}\, )\, ,
\end{equation}

\noindent
where the index $A$ ranges over all the generators of $G=G_{\rm bos}\times
U(1)_{R}$, so we have now

\begin{equation}
\label{eq:gaugeparameterselmag}
\alpha^{A}(x)\equiv \Lambda^{M}(x)\vartheta_{M}{}^{A}\, ,  
\end{equation}

\noindent
and the gauge transformations of the complex scalars, for instance, take the form

\begin{equation}
\label{eq:deltazelmag}
\delta Z^{i} = \Lambda^{M}\vartheta_{M}{}^{A}k_{A}{}^{i}\, .  
\end{equation}

The embedding tensor describes the embedding of the gauge group into the global symmetry group $G$. The part of the global symmetry group that cannot be embedded into $Sp(2n_{\rm V},\mathbb{R})$ is irrelevant for the purpose of gauging. There is thus no loss in generality to replace the global symmetry group by $Sp(2n_{\rm V},\mathbb{R})$. In this sense the embedding tensor provides an embedding of the gauge group into $Sp(2n_{\rm V},\mathbb{R})$. Besides the embedding into $Sp(2n_{V},\mathbb{R})$ there are further constraints that decrease the rank of the group
that we can actually gauge.

For instance, we must impose the constraint

\begin{equation}
\label{eq:quadraticE}
Q^{AB}\equiv {\textstyle\frac{1}{4}}\vartheta^{[A\vert M} \vartheta^{\vert B]}{}_{M} =0\,
,\,\,\,\,\Rightarrow\,\,\,\,  
\vartheta^{AM} \vartheta_{M}{}^{B}=0\, ,
\end{equation}

\noindent
which guarantees that the electric and magnetic gaugings are mutually local
\cite{deWit:2005ub} and we can go to a theory with only purely electric gaugings
by a symplectic transformation.

The embedding tensor must satisfy further conditions. We define the matrices

\begin{equation}
\label{eq:Xdef}
X_{MN}{}^{P} \equiv \vartheta_{M}{}^{A}T_{A\, N}{}^{P}\, ,
\end{equation}

\noindent which satisfy

\begin{equation}
X_{MNP}=X_{MPN}\, ,  
\end{equation}

\noindent
on account of Eq.~(\ref{eq:Tsymplectic}). Observe that the components
$\vartheta_{M}{}^{\sharp}$ are not present in the $X_{MNP}$ tensors. Further, we
impose the quadratic constraint\footnote{Observe that $\vartheta_{M}{}^{\sharp}$
  does not occur in $Q_{NM}{}^{A}$ either.}

\begin{equation}
\label{eq:quadraticTdef}
Q_{NM}{}^{A}
\equiv 
\vartheta_{N}{}^{A}T_{A\,  M}{}^{P}\vartheta_{P}{}^{A}
-\vartheta_{N}{}^{A}\vartheta_{M}{}^{B}f_{AB}{}^{A}=0\, ,  
\end{equation}

\noindent
to ensure invariance of $\vartheta_{M}{}^{A}$ and the \textit{representation
  constraint} \cite{deWit:2005ub}

\begin{equation}
\label{eq:linear}
L_{MNP}\equiv X_{(MNP)} = X_{(MN}{}^{Q}\Omega_{P)Q}=0\, .
\end{equation}

\noindent
This constraint is required by gauge invariance and supersymmetry\footnote{In
  Ref.~\cite{DeRydt:2008hw} it has been shown how this constraint gets modified
 in the presence of anomalies and the modifications can cancel exactly the lack of
  gauge invariance of the classical action.}. It
implies Eq.~(\ref{eq:XLSO}) and also

\begin{equation}
\label{eq:symmproperty}
X_{(MN)P}= -{\textstyle\frac{1}{2}}X_{PMN}\,\,\,
\Rightarrow
X_{(MN)}{}^{P} = Z^{PA}T_{AMN}\, ,
\end{equation}

\noindent
where we have defined

\begin{equation}
Z^{PA} \equiv
-{\textstyle\frac{1}{2}}\Omega^{NP}\vartheta_{N}{}^{A}\, .
\end{equation}

\noindent
This definition and that of the other projectors that appear in the
4-dimensional hierarchy are collected in Appendix~\ref{app-projectors}. The
tensor $Z^{PA}$ will be used to project in directions orthogonal to the
embedding tensor since, due to the first quadratic constraint
Eq.~(\ref{eq:quadraticE}),

\begin{equation}
Z^{MA}\vartheta_{M}{}^{B}= 0\, .
\end{equation}

Finally, it should be clear that the constraint
Eq.~(\ref{eq:superpotentialconstraint2}) on the triple product of embedding
tensor, momentum maps and superpotential should be generalized to

\begin{equation}
\label{eq:superpotentialconstraint3}  
(\vartheta_{M}{}^{\underline{\rm a}}\mathcal{P}_{\underline{\rm a}} 
+  
\vartheta_{M}{}^{\sharp}\mathcal{P}_{\sharp})\mathcal{L}=0\, .
\end{equation}

Regarding the gauging of the $U(1)_R$ symmetry group we have the following possibilities. If $\mathcal{L}=0$ then the gauging shows up in the covariant derivatives of the fermions through terms containing $\mathcal{P}_\sharp$. The covariant derivatives acting on the scalars and vectors do not `see' this gauging because $K_{\sharp}=T_{\sharp}=0$. If we have a non-vanishing superpotential then it must be that $\vartheta_M{}^{\underline{a}}\mathcal{P}_{\underline{a}}+\vartheta_M{}^{\sharp}\mathcal{P}_{\sharp}=0$ and in order to gauge the $U(1)_R$ symmetry it must be identified with a $U(1)$ subgroup of $G_{\rm{bos}}$.

We define gauge-covariant derivatives of objects
transforming according to $\delta\phi=\Lambda^{M}\delta_{M}\phi$ by 

\begin{equation}
\mathfrak{D}\phi = d\phi +A^{M}\delta_{M}\phi\, .  
\end{equation}

\noindent
The gauge fields transform according to

\begin{equation}
\label{eq:Astandardgaugetrans}
\delta A^{M} = -\mathfrak{D}\Lambda^{M} +\Delta A^{M}
= -(d\Lambda^{M}+X_{NP}{}^{M}A^{N}\Lambda^{P}) +\Delta A^{M}\, ,
\end{equation}

\noindent
where $\Delta A^{M}$ is a piece that we can add to this gauge transformation
if it satisfies

\begin{equation}
\vartheta_{M}{}^{A}\Delta A^{M}=0\, .  
\end{equation}

The covariant derivatives of the scalars, gravitino and chiralinos read

\begin{eqnarray}
\label{eq:DZ}
\mathfrak{D}Z^{i} & = & dZ^{i}
+A^{M}\vartheta_{M}{}^{A}k_{A}{}^{i}\, ,  \\
& & \nonumber \\
\mathfrak{D}_{\mu}\psi_{\nu} 
& = & 
\{\mathcal{D}_{\mu}-{\textstyle\frac{i}{2}}A^{M}{}_{\mu}\vartheta_{M}{}^{A}
\mathcal{P}_{A}\}\psi_{\nu}\, ,\\
& & \nonumber \\
\mathfrak{D}\chi^{i} 
& = & 
\mathcal{D}\chi^{i} +\Gamma_{jk}{}^{i}\mathfrak{D}Z^{j} \chi^{k}
-A^{M}\vartheta_{M}{}^{A}\partial_{j}k_{a}{}^{i}\chi^{j}
+{\textstyle\frac{i}{2}}A^{M}\vartheta_{M}{}^{A}
\mathcal{P}_{A}\chi^{i} \, .
\end{eqnarray}

Observe that $\Delta A^{M}$ drops automatically from the gauge transformations
of these expressions because $A^{M}$ always comes projected by
$\vartheta_{M}{}^{A}$. 

It is clear that we need to introduce auxiliary ``magnetic gauginos''
$\lambda_{\Lambda}$ in order to construct a symplectic vector of gauginos
$\lambda^{M}$ whose covariant derivative is

\begin{equation}
\mathfrak{D}\lambda^{M} 
= 
\{\mathcal{D}-{\textstyle\frac{i}{2}}A^{N}\vartheta_{N}{}^{A}\mathcal{P}_{A}\}\lambda^{M}
-X_{NP}{}^{M}A^{N}\lambda^{P}\, .  
\end{equation}

\noindent
The magnetic gauginos are the supersymmetric partners of the magnetic
1-forms. We will discuss their supersymmetry transformation rules later.

So far, to introduce the general 4-dimensional embedding-tensor formalism we
have introduced magnetic 1-forms $A_{\Lambda}$ and gauginos
$\lambda_{\Lambda}$. As discussed at the beginning of this section, we have to
find supersymmetry transformations for them and check the closure of the local
$N=1,d=4$ supersymmetry algebra.


\subsection{The supersymmetric hierarchy}
\label{sec:susy}

Before we deal with the supersymmetry transformations of the magnetic 1-forms
that we have introduced, we take one step back and study the closure of the
local $N=1,d=4$ supersymmetry algebra on the 0-forms.


\subsubsection{The scalars $Z^{i}$}

Their supersymmetry transformations are given by Eq.~(\ref{eq:susytransZN1}),
which we rewrite here for convenience:

\begin{equation}
\delta_{\epsilon} Z^{i} =
{\textstyle\frac{1}{4}} \bar{\chi}^{i}\epsilon\, .  
\end{equation}

\noindent
At leading order in fermions,

\begin{equation}
\delta_{\eta}\delta_{\epsilon}Z^{i}
=
{\textstyle\frac{1}{4}} \overline{(\delta_{\eta}\chi^{i})}\epsilon\, ,  
\end{equation}

\noindent
and all we need is the supersymmetry transformation for $\chi^{i}$. This is
given in Eq.~(\ref{eq:dilasusyruleN1}), which we also rewrite here

\begin{equation}
\delta_{\eta}\chi^{i} =
i\not\!\!\mathfrak{D} Z^{i}\eta^{*}
+2\mathcal{G}^{ij^{*}}\mathcal{D}_{j^{*}}\mathcal{L}^{*}\eta\, ,
\end{equation}

\noindent
where we have to take into account that the covariant derivative $\mathfrak{D}
Z^{i}$ is now given by Eq.~(\ref{eq:DZ}). We get

\begin{equation}
[\delta_{\eta}\, ,\, \delta_{\epsilon}]Z^{i}
=
\delta_{\mathrm{g.c.t.}}Z^{i}+\delta_{h} Z^{i}\, ,
\end{equation}

\noindent
where $\delta_{\mathrm{g.c.t.}}Z^{i}$ is a g.c.t.~with infinitesimal parameter
$\xi^{\mu}$

\begin{eqnarray}
\delta_{\mathrm{g.c.t.}}Z^{i} & = & \pounds_{\xi}Z^{i}=
+\xi^{\mu}\partial_{\mu} Z^{i}\, , \\
& & \nonumber \\
\label{eq:xi}
\xi^{\mu} & \equiv & 
\tfrac{i}{4}(\bar{\epsilon} \gamma^{\mu}\eta^{*}
-\bar{\eta}\gamma^{\mu}\epsilon^{*})\, ,
\end{eqnarray}

\noindent
and where $\delta_{h} Z^{i}$ is the gauge transformation
Eq.~(\ref{eq:deltazelmag}) with gauge parameter $\Lambda^{M}$

\begin{eqnarray}
\delta Z^{i} & = & \Lambda^{M}\vartheta_{M}{}^{A}k_{A}{}^{i}\, ,\\
& & \nonumber \\
\label{eq:LM}
\Lambda^{M} & \equiv & \xi^{\mu} A^{M}{}_{\mu}\, .
\end{eqnarray}

This is just a small generalization of the standard result in which electric
and magnetic gauge parameters appear. As expected, no duality relations are
required to close the local supersymmetry algebra on the $Z^{i}$.


\subsubsection{The 1-form fields  $A^{M}$}

As we have mentioned before, to define supersymmetry transformations for the
magnetic vectors $A_{\Lambda}$ it is convenient to introduce simultaneously
magnetic gauginos\footnote{Magnetic gauginos have also been introduced in
  Ref.~\cite{de Vroome:2007zd}.} $\lambda_{\Lambda}$. This is equivalent to
introducing $n_{V}$ auxiliary vector supermultiplets.  Symplectic covariance
suggests that we can write  the following supersymmetry transformation rules
for the electric and magnetic 1-forms and gauginos:

\begin{eqnarray}
\label{eq:susytransAMN1}
\delta_{\epsilon} A^{M}{}_{\mu} 
& = & 
-{\textstyle\frac{i}{8}}\bar\epsilon^{*}\gamma_{\mu}\lambda^{M} +\mathrm{c.c.}\, ,
\\
& & \nonumber \\
\label{eq:gaugsusyruleMN1}
\delta_{\epsilon}\lambda^{M} 
& = & 
{\textstyle\frac{1}{2}}\left[\not\!
F^{M+}+i \mathcal{D}^{M}\right]\epsilon\, , 
\end{eqnarray}

\noindent
where $F^{M}$ is the gauge-covariant 2-form field strength of $A^{M}$, to be
defined shortly, and where we have defined the symplectic vector 

\begin{equation}
\label{eq:DMdef}
\mathcal{D}^{M} \equiv
\left(
\begin{array}{c}
 \mathcal{D}^{\Lambda} \\  \mathcal{D}_{\Lambda} \\ 
\end{array}
\right)
\equiv
\left(
\begin{array}{c}
 \mathcal{D}^{\Lambda} \\ f_{\Lambda\Sigma} \mathcal{D}^{\Sigma}  \\ 
\end{array}
\right)\, ,
\end{equation}

\noindent
where now, the electric $\mathcal{D}^{\Lambda}$ has been redefined, with
respect to the purely electric gauging case, to include a term with the
magnetic component of the embedding tensor $\vartheta^{\Lambda A}$:

\begin{equation}
\label{eq:DLdef}
\mathcal{D}^{\Lambda} = - \Im {\rm m}f^{\Lambda\Sigma}\, (\vartheta_{\Sigma}{}^{A}
+f^{*}_{\Sigma\Omega}\vartheta^{\Omega A})\mathcal{P}_{A}\, .  
\end{equation}

Although at this point we do not need it, it is important to observe that
there is a duality relation between the magnetic gauginos and the electric
ones

\begin{equation}
\label{eq:dualL}
\lambda_{\Lambda}=f_{\Lambda\Sigma}\lambda^{\Sigma}\, .  
\end{equation}

\noindent
The gaugino duality relation is local and takes the same form as the duality
relation between the magnetic and the electric vector field strengths:

\begin{equation}
\label{eq:dualF}
F_{\Lambda}{}^{+} = f_{\Lambda\Sigma} F^{\Sigma\, +}\, ,
\end{equation}

\noindent
which is obtained from the duality between electric and magnetic vectors
$F_{\Lambda}=G_{\Lambda}$, combined with Eq.~(\ref{eq:MaxwellGl}). These
duality relations relate the supersymmetry transformation
$\delta_{\epsilon}\lambda^{\Lambda}$ to
$\delta_{\epsilon}\lambda_{\Lambda}$.

Now we can check the closure of the local supersymmetry algebra on $A^{M}$. It
is, however, convenient to know beforehand the form of the gauge transformations that we
should expect on the right hand side of the commutator. The gauge transformations of $A^{M}$
are given in Eq.~(\ref{eq:Astandardgaugetrans}) up to a term $\Delta A^{M}$
which is determined in the construction of the gauge-covariant field strength
$F^{M}$. This term is also needed to have well-defined supersymmetry
transformations for all the gauginos.

As shown in Ref.~\cite{deWit:2005ub}, this requires the
introduction of a set of 2-forms $B_{A}$ in $F^{M}$, which takes the form

\begin{equation}
\label{eq:FM}
F^{M} 
=  
dA^{M} +{\textstyle\frac{1}{2}}X_{[NP]}{}^{M}A^{N}\wedge A^{P} +Z^{MA}B_{A}\, ,  
\end{equation}

\noindent
and is gauge-covariant under the transformations\footnote{The label $h$ in the gauge transformations indicates that these are the gauge transformations as predicted by the tensor hierarchy.}

\begin{eqnarray}
\label{eq:deltahAMtext}
\delta_{h} A^{M} & = & -\mathfrak{D}\Lambda^{M} -Z^{MA}\Lambda_{A}\, ,
\\
& & \nonumber \\
\label{eq:deltahBAtext}
\delta_{h} B_{A} & = &  \mathfrak{D}\Lambda_{A} +2T_{A\, NP}[\Lambda^{N}F^{P}
+{\textstyle\frac{1}{2}}A^{N}\wedge\delta_{h} A^{P}] 
+\Delta B_{A}\, ,
\hspace{.5cm}  
Z^{MA}\Delta B_{A}=0\, .
\end{eqnarray}

Let us now compute the commutator of two supersymmetry transformations on
$A^{M}$. To leading order in fermions, Eq.~(\ref{eq:susytransAMN1}) gives

\begin{equation}
\delta_{\eta}\delta_{\epsilon}A^{M}{}_\mu
=
-{\textstyle\frac{i}{8}}
\bar{\epsilon}^{*}\gamma_{\mu}\delta_{\eta}\lambda^{M}
+\mathrm{c.c.}
\end{equation}

\noindent
Using Eq.~(\ref{eq:gaugsusyruleMN1}) with the parameter $\eta$, we find

\begin{equation}
[\delta_{\eta}\, ,\, \delta_{\epsilon}]A^{M}{}_\mu
=
\xi^{\nu}F^{M}{}_{\nu\mu}+Z^{MA}\mathcal{P}_{A}\xi_{\mu}\, ,
\end{equation}

\noindent
where $\xi^{\mu}$ is given by Eq.~(\ref{eq:xi}) and we have used 

\begin{equation}
\Im{\rm m}\mathcal{D}^{M} = 2Z^{MA}\mathcal{P}_{A}\, ,
\end{equation}

\noindent
which follows from the definitions Eqs.~(\ref{eq:DMdef}), (\ref{eq:DLdef}) and
(\ref{eq:Ztensordef}). We always expect a general coordinate transformation on
the right hand side of the form

\begin{equation}
\delta_{\mathrm{g.c.t.}}A^{M}{}_{\mu} = \pounds_{\xi}A^{M}{}_{\mu} =
\xi^{\nu}\partial_{\nu} A^{M}{}_{\mu}+\partial_{\mu}\xi^{\nu}A^{M}{}_{\nu}\, . 
\end{equation}

\noindent
Using the explicit form of the field strength $F^{M}$ Eq.~(\ref{eq:FM}) we can
rewrite it as

\begin{equation}
\delta_{\mathrm{g.c.t.}}A^{M}{}_{\mu} = 
\xi^{\nu}F^{M}{}_{\nu\mu} +\mathfrak{D}_{\mu}(A^{M}{}_{\nu}\xi^{\nu})
+Z^{MA}[B_{A\mu\nu}\xi^{\nu} -T_{A\, NP}A^{N}{}_{\mu}A^{P}{}_{\nu}\xi^{\nu}]\, .
\end{equation}

Using this expression in the commutator and the definition Eq.~(\ref{eq:LM})
of the gauge parameter $\Lambda^{M}$, we arrive at

\begin{equation}
[\delta_{\eta}\, ,\, \delta_{\epsilon}]A^{M}
=
\delta_{\mathrm{g.c.t.}}A^{M}+\delta_{h} A^{M}\, ,
\end{equation}

\noindent
where, in complete agreement with the tensor hierarchy, $\delta_{h} A^{M}$ is
the gauge transformation in Eq.~(\ref{eq:deltahAMtext}) with the 1-form gauge
parameter $\Lambda_{A}$ given by

\begin{eqnarray}
\label{eq:LA}
\Lambda_{A} & \equiv & -T_{A\, MN} A^{N}\Lambda^{M}+b_{A}-\mathcal{P}_{A}\xi\, ,\\
& & \nonumber \\
b_{A\, \mu} & \equiv & B_{A\, \mu\nu}\xi^{\nu}\, .
\end{eqnarray}

Observe that no duality relation was needed to close the local
supersymmetry algebra on the magnetic vector fields. This result is a
consequence of using fully independent magnetic gauginos as the supersymmetric
partners of the magnetic vector fields, i.e.~transforming as
$\delta_{\epsilon}\lambda_{\Sigma}\sim \not\! F_{\Sigma}{}^{+}$ instead of
$\delta_{\epsilon}\lambda_{\Sigma}\sim \not\! G_{\Sigma}{}^{+}$. In the later
case we would have gotten additional $G_{\Sigma}-F_{\Sigma}$ terms to be
cancelled by using the duality relation.


\subsubsection{The 2-form fields  $B_{A}$}

In order to have a gauge-covariant field strength $F^{M}$ for the 1-forms we
have been forced to introduce a set of 2-forms $B_{A}$ and now we want to
study the consistency of this addition to the theory from the point of view of
supersymmetry and gauge invariance. We will first study the closure of the
supersymmetry algebra on the 2-forms $B_{A}$ without introducing its
supersymmetric partners and, later on, we will introduce the 2-forms as
components of linear supermultiplets. In the first case, the local $N=1,d=4$
supersymmetry algebra will close up to the use of duality relations while in
the second case it will close exactly.

It is useful to know beforehand what to expect on the right hand side of the
commutator of two supersymmetry transformations acting on the 2-forms $B_{A}$. The
gauge transformations of the 2-forms are given in Eq.~(\ref{eq:deltahBAtext})
up to a term $\Delta B_{A}$ which is constraint to satisfy $Z^{MA}\Delta
B_{A}=0$. In Ref.~(\cite{Bergshoeff:2009ph}) it was found that, in general, 

\begin{equation}
\Delta B_{A}  = -Y_{AM}{}^{C}\Lambda_{C}{}^{M}\, ,
\end{equation}

\noindent
for some 2-form parameters $\Lambda_{C}{}^{M}$.  $Y_{AM}{}^{C}$ is the
projector given in Eq.~(\ref{eq:Ytensordef}) and is annihilated by $Z^{NA}$ by
virtue of the quadratic constraint Eq.~(\ref{eq:quadraticTel}) (see
Eq.~(\ref{eq:ZY=0})), as required by the gauge-covariance of
$F^{M}$. In generic 4-dimensional theories $Y_{AM}{}^{C}$ is the only tensor that is annihilated by $Z^{NA}$. At this point we have to remind ourselves that in $N=1,d=4$
supergravity there is another constraint, given in
Eq.~(\ref{eq:superpotentialconstraint3}), that may lead to additional terms in the gauge transformation 
of the 2-forms since Eq.~(\ref{eq:superpotentialconstraint3}) can be written as $Z^{MA}(\delta_{A}{}^{\underline{\rm a}}\mathcal{P}_{\underline{\rm a}} 
+\delta_{A}{}^{\sharp}\mathcal{P}_{\sharp})\mathcal{L}=0$. To see if there are any such additional terms in the gauge transformations of the 2-forms we need to compute the
commutator of two supersymmetry transformations on $B_{A}$.

In any case, the generic tensor hierarchy prediction is that, with the gauge
transformations Eq.~(\ref{eq:deltahBA}), which we rewrite here 

\begin{equation}
\delta_{h} B_{A} =  \mathfrak{D}\Lambda_{A} +2T_{A\, NP}[\Lambda^{N}F^{P}
+{\textstyle\frac{1}{2}}A^{N}\wedge\delta_{h} A^{P}] 
-Y_{AM}{}^{C}\Lambda_{C}{}^{M}\, , 
\end{equation}

\noindent
the gauge-covariant field strength of $B_{A}$ is as given in Eq.~(\ref{eq:HA})

\begin{equation}
H_{A} = \mathfrak{D}B_{A}
+T_{A\, RS}A^{R}\wedge[dA^{S} +{\textstyle\frac{1}{3}}X_{NP}{}^{S}A^{N}\wedge
A^{P}] +Y_{AM}{}^{C}C_{C}{}^{M}\, ,
\end{equation}

\noindent
where $C_{C}{}^{M}$ is a 3-form whose gauge transformations are determined to be

\begin{equation}
\delta_{h} C_{C}{}^{M} 
 =  
\mathfrak{D}\Lambda_{C}{}^{M}
 -F^{M}\wedge\Lambda_{C}
 -\delta_{h} A^{M}\wedge B_{C} 
-{\textstyle\frac{1}{3}}T_{C\, NP} A^{M}\wedge A^{N} \wedge \delta_{h} A^{P}
+\Lambda^{M}H_{C}
+\Delta C_{C}{}^{M}\, , 
\end{equation}

\noindent
where 

\begin{equation}
Y_{AM}{}^{C} \Delta C_{C}{}^{M}=0\, . 
\end{equation}

We will next see that Eq.~(\ref{eq:superpotentialconstraint3}) leads to additional terms in the 2-form gauge transformation. Inspired by the results of Ref.~\cite{Bergshoeff:2007ij}, we found that, for
the 2-forms $B_{A}$, the supersymmetry transformation is given by

\begin{equation}
\label{eq:susyBA}
\delta_{\epsilon}B_{A\mu\nu}
=
\tfrac{1}{4}[\partial_{i}\mathcal{P}_{A}\bar{\epsilon}\gamma_{\mu\nu}\chi^{i} 
+\mathrm{c.c.}] 
+\tfrac{i}{2} [\mathcal{P}_{A}\bar{\epsilon}^{*}\gamma_{[\mu}\psi_{\nu]}-\mathrm{c.c.}]
+2T_{AMN}A^M{}_{[\mu}\delta_\epsilon A^N{}_{\nu]}\, .
\end{equation}

\noindent 
The commutator of two of these supersymmetry transformations closes up to a
duality relation to be described later on, a general coordinate transformation, and
a gauge transformation of the form

\begin{equation}
\label{Eq:BAgaugetrafo}  
\delta^{\prime}_{h} B_{A} 
=  
\delta_{h}B_{A} 
-(\delta_{A}{}^{\underline{\rm a}}\mathcal{P}_{\underline{\rm a}} 
+\delta_{A}{}^{\sharp}\mathcal{P}_{\sharp})\Lambda\, ,
\end{equation}

\noindent
where $\delta_{h}B_{A}$ is the standard hierarchy gauge transformation
Eq.~(\ref{eq:deltahBA}) and where the 2-form parameters $\Lambda$ and
$\Lambda_{C}{}^{M}$ are given by

\begin{eqnarray}
\label{eq:LCM}
\Lambda_{C}{}^{M}
& \equiv &
-\Lambda^{M} B_{C}-c_{C}{}^{M}
-\tfrac{1}{3} T_{CQP}\Lambda^{P} A^{M}\wedge A^{Q}\, ,
\\
& & \nonumber \\
\label{Lambda-in-spinors}
\Lambda 
& \equiv & 
-c+2\Re{\rm e}(\phi \mathcal{L})\, ,
\\
& & \nonumber \\
\phi_{\mu\nu} 
& \equiv & 
\bar{\epsilon}^{*}\gamma_{\mu\nu}\eta^{*}
=-\bar{\eta}^{*}\gamma_{\mu\nu}\epsilon^{*}\, ,
\\
& & \nonumber \\
c_{C}{}^{M}{}_{\mu\nu} 
& \equiv & 
C_{C}{}^{M}{}_{\mu\nu\rho}\xi^{\rho}\, ,
\\
& & \nonumber \\
c_{\mu\nu} & \equiv & C_{\mu\nu\rho}\xi^{\rho}\, .
\end{eqnarray}

\noindent
The parameters $\Lambda^{M}$ and $\Lambda_{A}$ are, again, given by
Eqs.~(\ref{eq:LM}) and (\ref{eq:LA}), respectively. We have introduced the
anticipated 3-form $C$ with the gauge transformation

\begin{equation}
\delta^{\prime}_{h}C = -d\Lambda\, ,  
\end{equation}

\noindent
to take care of the St\"uckelberg shift parameter $\Lambda$. Strictly speaking
we only need to introduce $C$ when $\mathcal{L}\neq 0$ in which case, according to the
constraint Eq.~(\ref{eq:superpotentialconstraint3}),
$(\vartheta_{M}{}^{\underline{\rm a}}\mathcal{P}_{\underline{\rm a}} +
\vartheta_{M}{}^{\sharp}\mathcal{P}_{\sharp})=0$. We can express this as a
``constraint''

\begin{equation}
\label{eq:Cconstraint}
  (\vartheta_{M}{}^{\underline{\rm a}}\mathcal{P}_{\underline{\rm a}} 
  +  
  \vartheta_{M}{}^{\sharp}\mathcal{P}_{\sharp})C=0\, ,
\end{equation}

\noindent
so 

\begin{equation}
\label{eq:Lconstraint}
(\vartheta_{M}{}^{\underline{\rm a}}\mathcal{P}_{\underline{\rm a}} 
+  
\vartheta_{M}{}^{\sharp}\mathcal{P}_{\sharp})\Lambda=0\, .
\end{equation}

\noindent
This constraint ensures that
$Z^{MA}\Delta B_{A}=0$ so that $F^{M}$ remains gauge-covariant under
$\delta^{\prime}_{h}B_{A}$.

The success of closing the supersymmetry algebra on the 2-forms, $B_A$, that is evaluating the commutator of two supersymmetry transformations (\ref{eq:susyBA}), and showing that it gives rise to local symmetries acting on $B_A$ requires the use of the duality relation

\begin{equation}
\label{eq:dualityHj}
H^\prime_{A} = -{\textstyle\frac{1}{2}}\star j_{A}\, ,
\end{equation}

\noindent
where 

\begin{equation}
\label{eq:NoethercurrentA}
j_{A} \equiv 2k^{*}_{A\, i}\mathfrak{D}Z^{i} +\mathrm{c.c.}\, ,
\end{equation}

\noindent
is the covariant Noether current 1-form and where the hierarchy gauge-covariant field strength $H_{A}$ given in
Eq.~(\ref{eq:HA}) has been modified to:

\begin{equation}
\label{eq:HprimeA}
H^{\prime}_{A} \equiv
H_{A}
-(\delta_{A}{}^{\underline{\rm a}}\mathcal{P}_{\underline{\rm a}} 
+\delta_{A}{}^{\sharp}\mathcal{P}_{\sharp})C\, .
\end{equation}

\noindent
The modified field strength $H^\prime_A$ transforms covariantly under the modified gauge transformations (\ref{Eq:BAgaugetrafo}).

The right hand side of the duality relation (\ref{eq:dualityHj}) vanishes for
$A=\underline{\rm a},\sharp$. For these cases we expect to have currents
bilinear in fermions which cannot appear at the order in fermions we are
working at.

The origin of the extra term in Eq.~(\ref{eq:HprimeA}) that is proportional to $(\delta_{A}{}^{\underline{\rm a}}\mathcal{P}_{\underline{\rm a}}+\delta_{A}{}^{\sharp}\mathcal{P}_{\sharp})$ can be traced back to the fact that 
the identity

\begin{equation}
\label{eq:nplzero}
\partial^{i^{*}} \mathcal{P}_{\mathbf{a}}\mathcal{D}_{i^{*}}\mathcal{L}^{*}
-\mathcal{P}_{\mathbf{a}} \mathcal{L}^{*}=0\, ,
\end{equation}

\noindent
which is crucial for closing the supersymmetry algebra for the case $A=\mathbf{a}$ (it leads to a cancellation of terms coming from the supersymmetry variation of
the first and second terms of Eq.~(\ref{eq:susyBA})) cannot be extended to the
cases $A=\underline{\rm a},\sharp$ in which we have introduced \textit{fake}
(vanishing) Killing vectors. 

The introduction of the 3-forms $C$ and $C_A{}^M$ into the result for the commutator $[\delta_\eta,\delta_\epsilon]B_{A\mu\nu}$ via the duality relation (\ref{eq:dualityHj}) was necessary in order to make the result gauge invariant. Ultimately, this is only allowed if one can show that the supersymmetry algebra can also be closed on the 3-forms $C$ and $C_A{}^M$. This will be shown to be the case later on.


\subsubsection{The supermultiplet of $B_{A}$}

We are now going to show that if we add to the tensor hierarchy full linear
multiplets\footnote{Similar supermultiplets have been introduced in
  electro-magnetically gauged globally supersymmetric $N=2, d=4$ field theory
  \cite{de Vroome:2007zd}.} $\{B_{A\, \mu\nu},\varphi_{A},\zeta_{A}\}$ where
$\varphi_{A}$ is a real scalar and $\zeta_{A}$ is a Weyl spinor, instead of
just the 2-forms $B_{A}$, as in the preceding section, we can close the local
$N=1,d=4$ supersymmetry algebra on the 2-forms exactly without the use of the
duality relation Eq.~(\ref{eq:dualityHj}).

We will construct the supersymmetry rules of the linear supermultiplet first
for the case $A=\mathbf{a}$ after which this result will be generalized to
include also the cases $A=\underline{a},\sharp$. The above supersymmetry
transformation rule Eq.~(\ref{eq:susyBA}) suggests the fermionic duality rule

\begin{equation}
\label{eq:dualzA}
\zeta_{\mathbf{a}} = \partial_{i}\mathcal{P}_{\mathbf{a}}\chi^{i}= 
ik^{*}_{\mathbf{a}\, i}\chi^{i}\, ,  
\end{equation}

\noindent
so we would have 

\begin{equation}
\label{eq:susyBA2}
\delta_{\epsilon}B_{\mathbf{a}\mu\nu}
=
\tfrac{1}{4}[\bar{\epsilon}\gamma_{\mu\nu}\zeta_{\mathbf{a}} 
+\mathrm{c.c.}] 
+\tfrac{i}{2}
[\mathcal{P}_{\mathbf{a}}\bar{\epsilon}^{*}\gamma_{[\mu}\psi_{\nu]}
-\mathrm{c.c.}]
+2T_{\mathbf{a}\, MN}A^{M}{}_{[\mu}\delta_{\epsilon} A^{N}{}_{\nu]}\, .
\end{equation}

\noindent
The supersymmetry transformation rule of $\zeta_{\mathbf{a}}$ follows from the above
duality rule:

\begin{equation}
\delta_{\epsilon}\zeta_{\mathbf{a}} =  
ik^{*}_{\mathbf{a}\, i}\delta_{\epsilon}\chi^{i}
=-k^{*}_{\mathbf{a}\, i}\not\!\! \mathfrak{D}Z^{i} \epsilon^{*}
+2\partial_{i}\mathcal{P}_{\mathbf{a}}\mathcal{G}^{ij^{*}}
\mathcal{D}_{j^{*}}\mathcal{L}^{*}\epsilon \, .   
\end{equation}

\noindent
Using next the duality rule Eq.~(\ref{eq:dualityHj}) $j_{\mathbf{a}} =
4\Re{\rm e}(k^{*}_{\mathbf{a}\, i}\mathfrak{D}Z^{i}) =-2\star H_{\mathbf{a}}$
we find

\begin{equation}
\delta_{\epsilon}\zeta_{\mathbf{a}} 
=  
-i[{\textstyle\frac{i}{12}}\not\!\! H_{\mathbf{a}}+\Im{\rm m}(k^{*}_{\mathbf{a}\,
  i}\mathfrak{D}_{\mu}Z^{i} )\gamma^{\mu}] \epsilon^{*}
+2\mathcal{P}_{\mathbf{a}}\mathcal{L}^{*}\epsilon \, .
\end{equation}

To make contact with the standard linear multiplet supersymmetry
transformations we should be able to identify consistently

\begin{equation}
\Im{\rm m}(k^{*}_{\mathbf{a}\,
  i}\mathfrak{D}Z^{i} ) \equiv \mathfrak{D}\varphi_{\mathbf{a}}\, ,  
\end{equation}

\noindent
for some real scalar $\varphi_{\mathbf{a}}$. The integrability condition of this
equation can be obtained by acting with $\mathfrak{D}$ on both sides. Using
on the l.h.s.~the property

\begin{equation}
\mathfrak{D}k^{*}_{\mathbf{a}\, i} = 
\mathfrak{D}Z^{*j^{*}}\nabla_{j^{*}}k^{*}_{\mathbf{a}\, i}\, ,
\end{equation}

\noindent
and the Killing property, the integrability condition takes the form

\begin{equation}
-iF^{M}\vartheta_{M}{}^{\mathbf{b}}k^{*}_{[\mathbf{a}|i}k_{|\mathbf{b}]}{}^{i}= 
f_{\mathbf{ab}}{}^{\mathbf{c}}F^{M}\vartheta_{M}{}^{\mathbf{b}}\varphi_{\mathbf{c}}\, ,  
\end{equation}

\noindent
which is solved by 

\begin{equation}
-i k^{*}_{[\mathbf{a}|i}k_{|\mathbf{b}]}{}^{i}
= 
f_{\mathbf{ab}}{}^{\mathbf{c}}\varphi_{\mathbf{c}}\, .
\end{equation}

\noindent
Given that the Killing vectors can be derived from the Killing prepotential
$\mathcal{P}_{\mathbf{a}}$ which is equivariant, it follows that 

\begin{equation}
k^{*}_{[\mathbf{a}|i}k_{|\mathbf{b}]}{}^{i} 
= 
\tfrac{i}{2}\pounds_{\mathbf{a}}\mathcal{P}_{\mathbf{b}} 
= 
-\tfrac{i}{2} f_{\mathbf{ab}}{}^{\mathbf{c}}\mathcal{P}_{\mathbf{c}}\, , 
\end{equation}

\noindent
and we can finally identify 

\begin{equation}
\Im{\rm m}(k^{*}_{\mathbf{a}\, i}\mathfrak{D}Z^{i} )=
-\tfrac{1}{2} \mathfrak{D} \mathcal{P}_{\mathbf{a}}\, .
\end{equation}

\noindent
The supersymmetry transformations of the linear multiplet $\{B_{\mathbf{a}\,
  \mu\nu},\varphi_{\mathbf{a}},\zeta_{\mathbf{a}}\}$ are given by

\begin{eqnarray}
\delta_{\epsilon}\zeta_{\mathbf{a}} 
& = &   
-i[{\textstyle\frac{1}{12}}\not\!\! H_{\mathbf{a}}
+\not\!\!\mathfrak{D} \varphi_{\mathbf{a}}] \epsilon^{*}
-4\varphi_{\mathbf{a}}\mathcal{L}^{*}\epsilon \, ,\\
& & \nonumber \\
\delta_{\epsilon}B_{\mathbf{a}\mu\nu}
& = & 
\tfrac{1}{4}[\bar{\epsilon}\gamma_{\mu\nu}\zeta_{\mathbf{a}} 
+\mathrm{c.c.}] 
-i[\varphi_{\mathbf{a}}\bar{\epsilon}^{*}\gamma_{[\mu}\psi_{\nu]}-\mathrm{c.c.}]
+2T_{\mathbf{a}\, MN}A^M{}_{[\mu}\delta_\epsilon A^N{}_{\nu]}\, , \\
& & \nonumber \\
\delta_{\epsilon}\varphi_{\mathbf{a}} 
& = & 
-\tfrac{1}{8}\bar{\zeta}_{\mathbf{a}}\epsilon +\mathrm{c.c.}\, .
\end{eqnarray}

The duality relations needed to relate these fields to the fundamental fields
of the $N=1,d=4$ gauged supergravity are

\begin{eqnarray}
\zeta_{\mathbf{a}} & = & \partial_{i}\mathcal{P}_{\mathbf{a}}\chi^{i}\, ,\label{eq:dualityLmuliplet1}\\
& & \nonumber \\
H_{\mathbf{a}} & = & -\tfrac{1}{2}\star j_{\mathbf{a}}\, ,\\
& & \nonumber \\
\varphi_{\mathbf{a}} & = &   -\tfrac{1}{2} \mathcal{P}_{\mathbf{a}}\, .\label{eq:dualityLmuliplet3}
\end{eqnarray}

The supersymmetry algebra closes on all the fields of the linear multiplet
without the use of any duality relation. 

Now that we know the supersymmetry transformation rules for $A=\mathbf{a}$ we
will generalize them to all values of $A$. The supersymmetry transformations
of the linear multiplet $\{B_{A\, \mu\nu},\varphi_{A},\zeta_{A}\}$ are given
by

\begin{eqnarray}
\label{eq:deltazetaA}
\delta_{\epsilon}\zeta_{A} 
& = &   
-i[{\textstyle\frac{1}{12}}\not\!\! H^{\prime}_{A}
+\not\!\!\mathfrak{D} \varphi_{A}] \epsilon^{*}
-4\delta_{A}{}^{\mathbf{a}}\varphi_{\mathbf{a}}\mathcal{L}^{*}\epsilon\, ,
\\
& & \nonumber \\
\delta_{\epsilon}B_{A\mu\nu}
& = & 
\tfrac{1}{4}[\bar{\epsilon}\gamma_{\mu\nu}\zeta_{A} 
+\mathrm{c.c.}] 
-i[\varphi_{A}\bar{\epsilon}^{*}\gamma_{[\mu}\psi_{\nu]}-\mathrm{c.c.}]
+2T_{A\, MN}A^{M}{}_{[\mu}\delta_{\epsilon} A^{N}{}_{\nu]}\, , \\
& & \nonumber \\
\delta_{\epsilon}\varphi_{A} 
& = & 
-\tfrac{1}{8}\bar{\zeta}_{A}\epsilon +\mathrm{c.c.}\, .
\end{eqnarray}

The duality relations, Eqs.~(\ref{eq:dualityLmuliplet1}) to (\ref{eq:dualityLmuliplet3}), become

\begin{eqnarray}
\zeta_{A} & = &  \partial_{i}\mathcal{P}_{A}\chi^{i}\, ,\\
& & \nonumber \\
H^{\prime}_{A} & = & -\tfrac{1}{2}\star j_{A}\, ,\\
& & \nonumber \\
\varphi_{A} & = &   -\tfrac{1}{2} \mathcal{P}_{A}\, .
\end{eqnarray}

\noindent
Observe that some terms on the right hand side are zero for $A=\underline{\rm
  a},\sharp$, at least to leading order in fermions. 

Now the gauge parameters that appear on the right hand side of the commutator
of two supersymmetry transformations are different from those we found in the
previous section and, therefore, do not match with those we found in the case
of the 1-forms. To relate the parameters of the supersymmetry algebra in the
case with and without the linear supermultiplets we also need to use the above
duality relations. For instance, $\Lambda_{A}$ is given by Eq.~(\ref{eq:LA})
with $\mathcal{P}_{A}$ replaced by $-2\varphi_{A}$. This means that, in order
to supersymmetrize consistently the tensor hierarchy we also must replace
$\mathcal{P}_{A}$ by $-2\varphi_{A}$ in the supersymmetry transformation rules
of the gauginos Eq.~(\ref{eq:gaugsusyruleMN1}) (i.e.~in the definition of
$\mathcal{D}^{M}$ Eqs.~(\ref{eq:DMdef}) and (\ref{eq:DLdef})).  There are
furthermore also 3-forms contained in the transformation rule for
$\zeta_A$. Thus, if we continue this program we need to find a way to close
the algebra on all the 3-forms without using any duality relations.

However, we will not pursue here any further the supersymmetrization of the
tensor hierarchy for the higher-rank $p$-forms but we think that the above
results strongly suggest that an extension with additional fermionic and
bosonic fields of the tensor hierarchy on which the local supersymmetry
algebra closes without the use of duality relations must exist. The duality
relations must project the supersymmetric tensor hierarchy on to the $N=1$
supersymmetric generalization of the (bosonic) action which will be given later in
Eq.~(\ref{eq:action1-2-forms}).

As we have seen in the vector and 2-form cases, the duality relations among
the additional fields (fermionic $\lambda_{\Sigma},\zeta^{A}$ and bosonic
$\varphi_{A}$) are local as opposed to those involving the original bosonic
fields ($A_{\Lambda}$, $B_A$), which are non-local and related via Hodge-duality.


\subsubsection{The 3-form fields  $C_{A}{}^{M}$}

We will be brief here because the construction of the field strength and the
determination of the gauge transformations of the 3-forms $C_{A}{}^{M}$ are
similar to those of the other fields. 

We first remark that, in order to make the standard hierarchy's field strength
$G_{C}{}^{M}$ gauge-invariant under the new gauge transformations, we must
modify it as follows:

\begin{equation}
G^{\prime}_{A}{}^{M} 
\equiv 
G_{A}{}^{M} 
+(\delta_{A}{}^{\underline{\rm a}}\mathcal{P}_{\underline{\rm a}} 
+\delta_{A}{}^{\sharp}\mathcal{P}_{\sharp}) D^{M}\, ,
\end{equation}

\noindent
where $G_{A}{}^{M}$ is given in Eq.~(\ref{eq:GCM}) and $D^{M}$ is a 4-form
transforming as

\begin{equation}
\delta^{\prime}_{h}D^{M}= 
\mathfrak{D}\Sigma^{M} 
+(F^{M}-{\textstyle\tfrac{1}{2}Z^{MA}B_{A}})\wedge \Lambda\, ,  
\end{equation}

\noindent
and where we must also modify the gauge transformation rules of the 3-forms
$C_{A}{}^{M}$ to be 

\begin{equation}
\delta^{\prime}_{h} C_{A}{}^{M} 
= 
\delta_{h} C_{A}{}^{M} -(\delta_{A}{}^{\underline{\rm a}}\mathcal{P}_{\underline{\rm a}} 
+\delta_{A}{}^{\sharp}\mathcal{P}_{\sharp}) \mathfrak{D}\Sigma^{M}\, .  
\end{equation}

In order to prove this result we have made use of the constraint
Eq.~(\ref{eq:superpotentialconstraint3}) and also of the fact, mentioned in
Section~\ref{sec-pertursymm}, that the directions $A=\underline{\rm a}$ for
which $\mathcal{P}_{\underline{\rm a}}\neq 0$ must necessarily be Abelian, so 

\begin{equation}
\label{eq:superpotentialconstraint4}  
Y_{AM}{}^{A}(\delta_{A}{}^{\underline{\rm a}}\mathcal{P}_{\underline{\rm a}} 
+  
\delta_{A}{}^{\sharp}\mathcal{P}_{\sharp})\mathcal{L}=0\, ,
\end{equation}

\noindent
etc.

Then, the supersymmetry transformations of the 3-forms $C_{A}{}^{M}$ are given
by

\begin{equation}
\label{eq:susyCAM}
\delta_{\epsilon}C_{A}{}^{M}{}_{\mu\nu\rho}
=
-\tfrac{i}{8} [\mathcal{P}_{A}\bar{\epsilon}^{*}\gamma_{\mu\nu\rho}\lambda^{M}
-\mathrm{c.c.}]
-3 B_{A\, [\mu\nu|}\delta_{\epsilon} A^{M}{}_{|\rho]}
-2T_{A\, PQ}A^{M}{}_{[\mu}A^{P}{}_{\nu|} \delta_{\epsilon} A^{Q}{}_{|\rho]}\, .
\end{equation}

The local $N=1,d=4$ supersymmetry algebra closes on $C_{A}{}^{M}$ upon the use
of a duality relation to be discussed later. The gauge transformations of
$C_{A}{}^{M}$ that appear on the right hand side are the ones described above
with 

\begin{eqnarray}
\Lambda_{BC}
& = &
d_{BC}
+B_{[B}\wedge b_{C]}
+2 T_{[B\vert\, NP}\Lambda^{P} A^{N}\wedge B_{C]}\, ,\\
& & \nonumber\\
\Lambda^{NPQ}
& = &
d^{NPQ}
+2\Lambda^{(P} A^{N}\wedge(F^{Q)}-Z^{Q)C}B_{C})
-\tfrac{1}{4}X_{RS}{}^{(Q}\Lambda^{P} A^{N)}\wedge A^{R}\wedge A^{S}\, ,\\
& & \nonumber\\
\Lambda_{E}{}^{NP}
& = &
d_{E}{}^{NP}
-\Lambda^{N} C_{E}{}^{P}
+\tfrac{1}{2}T_{E\, QR}\Lambda^{Q} A^{N}\wedge A^{R}\wedge A^{P}\, ,
\end{eqnarray}

\noindent
where $d_{BC\mu\nu\rho}=D_{BC\mu\nu\rho\sigma}\xi^{\sigma}$, and similarly for
$d^{NPQ}$ and $d_{E}{}^{NP}$. The gauge transformation parameters
$\Lambda^{M}$, $\Lambda_{\mathbf{a}}$ and $\Lambda_{\mathbf{a}}{}^{M}$ are,
again, given by Eqs.~(\ref{eq:LM}), (\ref{eq:LA}) and (\ref{eq:LCM}),
respectively.

In the closure of the local supersymmetry algebra we have made use of the
duality relation

\begin{equation}
G^{\prime}_{A}{}^{M}
=
-\tfrac{1}{2}\star\Re\rm{e}(\mathcal{P}_{A}\mathcal{D}^{M})\, . 
\end{equation}

\noindent
According to the results of Ref.~\cite{Bergshoeff:2009ph}, the duality
relation has the general form 

\begin{equation}
G^{\prime}_{A}{}^{M}
=
\tfrac{1}{2}\star\frac{\partial V}{\partial \vartheta_{M}{}^{A}}\, . 
\end{equation}

\noindent
Comparing these two expressions and using the relation between the potential
of the supergravity theory and the fermion shifts, we conclude that, after the
general electric-magnetic gauging the potential of $N=1,d=4$ supergravity is
given by   

\begin{equation}
\label{eq:Ve-mg}
V_{\rm e-mg} 
= 
V_{\rm u}
-\tfrac{1}{2} \Re{\rm e}\, \mathcal{D}^{M}\vartheta_{M}{}^{A} \mathcal{P}_{A}
=
V_{\rm u}
+\tfrac{1}{2} \mathcal{M}^{MN}\vartheta_{M}{}^{A}\vartheta_{N}{}^{A}
\mathcal{P}_{A} \mathcal{P}_{B} \, , 
\end{equation}

\noindent
where $\mathcal{M}$ is the symplectic matrix defined in
Eq.~(\ref{eq:Mdef}). It satisfies 

\begin{equation}
\label{eq:dVdt}
\partial V_{\rm e-mg}/\partial\vartheta_{M}{}^{A} = 
-\Re{\rm e}(\mathcal{D}^{M}\mathcal{P}_{A})\, . 
\end{equation}

There may exist a supermultiplet containing the 3-forms $C_{A}{}^{M}$ such
that the supersymmetry algebra closes without the need to use a duality
relation. We leave it to future work to study its possible (non-)existence.


\subsubsection{The 3-form $C$ and the dual of the superpotential}
\label{subsec:CandCprime}

We have seen that the consistency of the closure of the local supersymmetry
algebra on the 2-forms $B_{\underline{\rm a}}$ and $B_{\sharp}$ requires the
existence of a 3-form field that we have denoted by $C$, whose gauge
transformation cancels the St\"uckelberg shift of those 2-forms.

An Ansatz for the supersymmetry transformation of $C$ can be made by writing
down 3-form spinor bilinears that have zero K\"ahler weight and that are
consistent with the chirality of the fermionic fields. Further, from
Eq.~(\ref{Lambda-in-spinors}) it follows that there will be no gauge potential
terms needed in the Ansatz. We thus make the following Ansatz

\begin{equation}
\label{eq:3formansatz}
\delta_{\epsilon} C_{\mu\nu\rho}
= 
-3i\eta \mathcal{L} \,\bar{\epsilon}^{*} \gamma_{[\mu\nu}\psi^{*}_{\rho]}
-\tfrac{1}{2}\eta \mathcal{D}_{i} \mathcal{L} 
\bar{\epsilon}^{*}\gamma_{\mu\nu\rho}\chi^{i}+\mathrm{c.c.}\, ,
\end{equation}

\noindent
where $\eta$ is a constant to be found. It turns out that the local
supersymmetry algebra closes for two different reality conditions for $\eta$,
which leads to the existence of two different 3-forms that we will call $C$
and $C^{\prime}$.

\begin{enumerate}

\item For $\eta=-i$ the algebra closes into the gauge transformations required
  by the 2-forms $B_{\underline{\rm a}}$ and $B_{\sharp}$ provided that the
  field strength $G=dC$ vanishes. As discussed earlier there may be
  non-vanishing contributions if we were to construct the supersymmetry
  algebra at the quartic fermion order.
  
\item For $\eta\in\mathbb{R}$ the algebra closes into the following gauge
  transformation

\begin{equation}
\delta_{\mathrm{gauge}} C^{\prime}=-d\Lambda^{\prime}\, ,
\end{equation}

\noindent
where the 2-form $\Lambda^{\prime}$ is given by 

\begin{equation}
\Lambda^{\prime} =  c^{\prime}-2 \eta\Im\mathrm m(\mathcal L\phi)\, ,
\hspace{1cm}
c^{\prime}_{\mu\nu} \equiv C^{\prime}_{\mu\nu\rho}\xi^{\rho}\, ,
\end{equation}

\noindent
provided the field strength $G^{\prime}=dC^{\prime}$ satisfies the duality relation

\begin{equation}
G^{\prime}= \star \eta 
(-24\vert\mathcal L\vert^{2} 
+8\mathcal G^{ij^*}\mathcal{D}_{i} \mathcal{L}
\mathcal{D}_{j^{*}}\mathcal{L}^{*})\, .
\end{equation}

\end{enumerate}

Observe that the right hand side is nothing but the part of the scalar
potential Eq.~(\ref{eq:Ve-mg}) that depends on the superpotential. Actually,
if we rescale the superpotential by $\mathcal{L} \rightarrow \eta
\mathcal{L}$, then we can rewrite the above duality relation in the standard
fashion

\begin{equation}
G^{\prime}= \tfrac{1}{2}\star \frac{\partial V_{\rm e-mg}}{\partial \eta}\, ,
\end{equation}

\noindent
and, therefore, we can see the 3-form $C^{\prime}$ as the dual of the
deformation parameter associated to the superpotential, just as we can see the
3-forms $C_{A}{}^{M}$ as the duals of the deformation parameters
$\vartheta_{M}{}^{A}$.

Observe that, had we chosen to work with a vanishing superpotential we would
have found the duality rule $G^{\prime}=0$. This suggests a possible
interpretation of the 3-form $C$ to be explored: that it may be related to
another, as yet unknown, deformation of $N=1,d=4$ supergravity which has not
been used. The full supersymmetric action is needed to confirm this
possibility or to find, perhaps, a term bilinear in fermions which is dual to
$C$.

Finally, observe that neither of the 3-forms $C,C^{\prime}$ was predicted by
the standard tensor hierarchy. $C$, though, is predicted by the extension
associated to the constraints Eqs.~(\ref{eq:superpotentialconstraint3})
and (\ref{eq:superpotentialconstraint4}).


\subsubsection{The 4-form fields  $D_{E}{}^{NP},D_{AB},D^{NPQ}, D^{M}$}
\label{subsec:4forms}

In the previous sections we have introduced four 4-forms
$D_{E}{}^{NP},D_{AB},D^{NPQ},D^M$ in order to close the local supersymmetry
algebra and have fully gauge-covariant field strengths. We thus expect that we
can also find consistent supersymmetry transformations for all these 4-forms.

For the three 4-forms $D_{E}{}^{NP},D_{AB},D^{NPQ}$ there is a slight
complication that has to do with the existence of extra St\"uckelberg shift
symmetries. There are two such shift symmetries and in
Appendix~\ref{app-gaugetranshierarchy} they correspond to the parameters
$\tilde{\Lambda}_{E}{}^{(NP)}$ and $\Lambda_{BE}{}^{P}$. The origin of these
symmetries lies in the fact that the $W$ tensors that appear in the field
strengths of the 3-forms are not all independent.  The symmetries result from
the identities \ref{eq:Wrelation1} and \ref{eq:Wrelation2} together with the
constraints $L_{NPQ}=Q^{AB}=Q_{NM}{}^A=0$. This means that if we want to
realize $N=1$ supersymmetry on the 4-forms $D_{E}{}^{NP},D_{AB},D^{NPQ}$ the
parameters $\tilde{\Lambda}_{E}{}^{(NP)}$ and $\Lambda_{BE}{}^{P}$ will appear
on the right hand side of commutators as part of the local algebra.

Most of these features are already visible in the simpler case of the ungauged
theory\footnote{Note that the hierarchy remains non-trivial for
  $\vartheta_{M}{}^{A}=0$.}, i.e. for $\vartheta_{M}{}^{A}=0$ and even when
the ungauged case has no symmetries that act on the vectors, i.e.~when all the
matrices $T_{A}=0$. We will restrict ourselves to realizing the supersymmetry
algebra on the 4-forms for the ungauged theory with $T_{A}=0$ for all $A$ for
simplicity. The 4-form supersymmetry transformations in this simple setting
are given by

\begin{eqnarray}
\delta_{\epsilon} D_{AB}
& = &
-\tfrac{i}{2}  \star \mathcal{P}_{[A}\partial_{i} \mathcal{P}_{B]}\bar{\epsilon} 
\chi^{i}+\mathrm{c.c.}- B_{[A}\wedge\delta_{\epsilon} B_{B]}\, ,
\\
& & \nonumber \\
\delta_\epsilon D^{NPQ}
& = &
10 A^{(N}\wedge F^{P}\wedge \delta_{\epsilon} A^{Q)}\, ,
\\
& & \nonumber \\
\delta_{\epsilon} D_{E}{}^{NP}
& = &
C_{E}{}^{P}\wedge \delta_{\epsilon} A^{N}\, .\\
& & \nonumber \\
\delta_{\epsilon} D^{M} & = &
-\tfrac{i}{2}\star\mathcal L^*\bar\epsilon \lambda^M+\mathrm{c.c.}
+C\wedge\delta_{\epsilon}A^M\,.
\end{eqnarray}

\noindent
When $\vartheta_{M}{}^{A}=0$ and $T_{A}=0$ the only place where there still
appears a St\"uckelberg shift parameter is in the gauge transformation of
$D_{E}{}^{NP}$. From the commutators we find that

\begin{equation}
\tilde{\Lambda}_{E}{}^{(NP)}=-2\Lambda^{(N}F^{P)}\wedge B_{E}\, .
\end{equation}


\subsection{The gauge-invariant bosonic action}

It turns out that in order to write an action for the bosonic fields of the
theory with electric and magnetic gaugings of perturbative and
non-perturbative symmetries it is enough to add to the fundamental (electric)
fields just the magnetic 1-forms $A_{\Lambda}$ and the 2-forms $B_{A}$. The
gauge-invariant action takes the form

\begin{equation}
\label{eq:action1-2-forms}
\begin{array}{rcl}
S_{\rm e-mg}  & = &  {\displaystyle\int}
\left\{
\star R -2\mathcal{G}_{ij^{*}}\mathfrak{D}Z^{i} 
\wedge\star\mathfrak{D}Z^{*\, j^{*}}
-2\Im{\rm m}f_{\Lambda\Sigma} 
F^{\Lambda}\wedge \star F^{\Sigma}
+2\Re{\rm e}f_{\Lambda\Sigma} 
F^{\Lambda} \wedge F^{\Sigma}
\right.
\\
& & \\
& & 
-\star V_{\rm e-mg}
-4Z^{\Sigma A}B_{A}\wedge 
\left(F_{\Sigma} -{\textstyle\frac{1}{2}}Z_{\Sigma}{}^{B}B_{B}\right)
-{\textstyle\frac{4}{3}}  X_{[MN]\Sigma} A^{M}\wedge A^{N} \wedge
\left(F^{\Sigma} -Z^{\Sigma B}B_{B}\right)
\\
\\
& & \\
& & 
\left.
-{\textstyle\frac{2}{3}}  X_{[MN]}{}^{\Sigma}A^{M}\wedge A^{N} \wedge
\left(dA_{\Sigma} -{\textstyle\frac{1}{4}} X_{[PQ]\Sigma} A^{P}\wedge A^{Q}\right)
\right\}\, .\\
\end{array}
\end{equation}

\noindent
The scalar potential $V_{\rm e-mg}$ is given by Eq.~(\ref{eq:Ve-mg}).
Furthermore, the gauge transformations that leave invariant the above action
($\delta_{a}$) are those of the extended hierarchy ($\delta^{\prime}_{h}$)
except for the 2-forms:

\begin{equation}
\label{eq:actiongaugetransformations3} 
\delta_{a} B_{A} =  \delta^{\prime}_{h}B_{A} -2T_{A\,
  NP}\Lambda^{N}(F^{P}-G^{P})\, . 
\end{equation}

\noindent
The action contains the 2-forms $B_{A}$ always contracted with $Z^{MA}$ so
that we do not need to worry about the different behavior of $B_{\mathbf{a}}$
and $B_{\underline{\rm a}},B_{\sharp}$ under gauge transformation due to the
extra constraint Eq.~(\ref{eq:Lconstraint}).

A general variation of the above action gives

\begin{equation}
\label{eq:variationaction1-2-forms}
\delta S={\displaystyle\int}
\left\{
\delta g^{\mu\nu}  {\displaystyle\frac{\delta S}{\delta g^{\mu\nu}}}
+\left(\delta Z^{i}  {\displaystyle\frac{\delta S}{\delta Z^{i}}} 
+\mathrm{c.c.}\right)
-\delta A^{M}\wedge \star {\displaystyle\frac{\delta S}{\delta A^{M}}}
+2\delta B_{A} \wedge \star {\displaystyle\frac{\delta S}{\delta  B_{A}}}
\right\}\, ,
\end{equation}

\noindent
where the first variations with respect to the different fields are given by

\begin{eqnarray}
\label{eq:SEmn}
-\star{\displaystyle\frac{\delta S}{\delta g^{\mu\nu}}} & = & 
G_{\mu\nu}
+2\mathcal{G}_{ij^{*}}[\mathfrak{D}_{\mu}Z^{i} \mathfrak{D}_{\nu}Z^{*\, j^{*}}
-{\textstyle\frac{1}{2}}g_{\mu\nu}
\mathfrak{D}_{\rho}Z^{i}\mathfrak{D}^{\rho}Z^{*\, j^{*}}]
\nonumber \\
& & \nonumber \\
& & 
-G^{M}{}_{(\mu|}{}^{\rho}\star G_{M| \nu) \rho}
+{\textstyle\frac{1}{2}}g_{\mu\nu}V_{\rm e-mg}\, ,
\\
& & \nonumber \\
\label{eq:SEi}
-{\textstyle \frac{1}{2}} {\displaystyle\frac{\delta S}{\delta Z^{i}}} & = & 
\mathcal{G}_{ij^{*}}\mathfrak{D}\star \mathfrak{D} Z^{*\, j^{*}}
-\partial_{i}G_{M}{}^{+}\wedge G^{M +}
-\star {\textstyle\frac{1}{2}}\partial_{i}V_{\rm e-mg}\, ,
\\
& & \nonumber \\
-{\textstyle\frac{1}{4}}{\displaystyle\star \frac{\delta S}{\delta  A^{M}}}
& = &
\mathfrak{D}G_{M}-{\textstyle\frac{1}{4}}
\vartheta_{M}{}^{A}\star j_{A} 
+{\textstyle\frac{1}{2}}T_{A\, MN}A^{N}\wedge \vartheta^{PA}(F_{P}-G_{P})\, , \\
& & \nonumber \\
\label{eq:SEA}
{\displaystyle\star \frac{\delta S}{\delta  B_{A}}}
& = & 
\vartheta^{PA}(F_{P}-G_{P})\, .
\end{eqnarray}

The above equations are formally symplectic-covariant and, therefore,
electric-magnetic duality symmetric. Both the Maxwell equations and the
``Bianchi identities'' have now sources to which they couple with a strength
determined by the embedding tensor's electric and magnetic components.

It is expected to be possible to find a gauge-invariant action in which all
the hierarchy's fields appear (as was done in \cite{Bergshoeff:2009ph}) if one
assumes that none of the constraints on the embedding tensor are
satisfied. Then, the 3-forms $C_{A}{}^{M}$ and the 4-forms
$D_{E}{}^{NP},D_{AB},D^{NPQ}$, $D^M$ are introduced as Lagrange multipliers
enforcing the constancy of the embedding tensor and the algebraic constraints
$Q_{NP}{}^{E}=0,Q^{AB}=0$, $L_{NPQ}=0$ and $(\vartheta_{M}{}^{\underline{\rm
    a}}\mathcal{P}_{\underline{\rm
    a}}+\vartheta_{M}{}^{\sharp}\mathcal{P}_{\sharp})\mathcal{L}=0$,
respectively, but we will not study this possibility here.

It should be stressed that, even though the action
Eq.~(\ref{eq:action1-2-forms}) contains $2n_{V}$ vectors and some number
$n_{B}$ of 2-forms $B_{a}$ it does not carry all those degrees of freedom. To
make manifest the actual number of degrees of freedom we briefly repeat here
the arguments of \cite{deWit:2005ub} regarding the gauge fixing of the action
(\ref{eq:action1-2-forms}). First, we choose a basis of magnetic vectors and
generators such that the non-zero entries of $\vartheta^{\Lambda a}$ arrange
themselves into a square invertible submatrix $\vartheta^{Ii}$. We split
accordingly $A_{\Lambda\mu}=(A_{I\mu},A_{U\mu})$. It can be shown by looking
at the vector equations of motion that the Lagrangian does not depend on the
$A_{U\mu}$, i.e.~$\delta\mathcal{L}/\delta A_{U\mu}=0$. Further, the electric
vectors $A^{I}{}_{\mu}$ that are dual to the magnetic vectors $A_{I\mu}$,
which are used in some gauging, have massive gauge transformations, $\delta
A^{I}{}_{\mu}=-\mathfrak{D}_{\mu}\Lambda^{I}-\vartheta^{Ii}\Lambda_{i\mu}$ and
can be gauged away. The $n_B$ 2-forms $B_{i}$ can by eliminated from the
Lagrangian by using their equations of motion Eq.~(\ref{eq:SEA}). The 2-forms
appear without derivatives in Eq.~(\ref{eq:SEA}) so that it is possible to
solve for them and to substitute the on-shell expression back into the
action. This is allowed as the 2-forms appear everywhere (up to partial
integrations) without derivatives. One then ends up with an action depending
on $n_{B}$ magnetic vectors $A_{I\mu}$ and $n_{V}-n_{B}$ electric vectors
$A^{U}{}_{\mu}$.

The relation between the tensor hierarchy and the action (or its equations of
motion) as well as the physical interpretation of the field content of the
extended hierarchy will be discussed in the next section.


\section{Summary and conclusions}
\label{sec-conclusions}

We have discussed the possible symmetries of $N=1, d=4$ supergravity and their
gauging using as gauge fields both electric and magnetic vectors. 

When using both electric and magnetic 1-forms as gauge fields at the same time
one is also compelled to introduce 2-forms $B_{A}$, associated to all the
possible symmetries of the theory. For each electric vector $A^{\Lambda}$
whose magnetic dual $A_{\Lambda}$ is gauged, because the magnetic components of
the embedding tensor $\vartheta^{\Lambda A}$ do not vanish, one introduces a
2-form $\vartheta^{\Lambda A}B_{A}$ in its field strength.  $A^{\Lambda}$ has
a massive gauge transformation and it forms a St\"uckelberg pair with the
2-form $\vartheta^{\Lambda A}B_{A}$. By electro-magnetic duality we end up
with St\"uckelberg pairs $A^{M},\vartheta_{M}{}^{A}B_{A}$.

The embedding tensor-projected 2-forms $\vartheta_{M}{}^{\bf a}B_{\bf a}$ are
dual to the embedding tensor-projected Noether currents that are associated to
gauged isometry directions $\vartheta_{M}{}^{\bf a}j_{\bf a}$ whereas the
remaining 2-forms $B_{\bf a}$ are dual to ungauged isometry directions. The
2-forms $B_{\underline{a}}$ and $B_{\sharp}$ are pure gauge at lowest order in
fermions, but it is to be expected that they are actually dual to the Noether
currents associated to the respective symmetries, which are bilinear in
fermions. To properly test this idea one would have to construct the
supersymmetry algebra at quartic order in fermions.

We have seen that the presence of a non-vanishing superpotential breaks the
global symmetries that we have denoted with the indices > $\underline{\rm
  a},\#$. Thus, if $\mathcal{L}\neq 0$, we must set
$(\vartheta_{M}{}^{\underline{\rm a}}\mathcal{P}_{\underline{\rm a}} +
\vartheta_{M}{}^{\sharp}\mathcal{P}_{\sharp})=0$, which is a new constraint
that the embedding tensor must satisfy. We have written it in the form
Eq.~(\ref{eq:superpotentialconstraint3}) to handle the cases $\mathcal{L}=0$
and $\mathcal{L}\neq 0$ simultaneously. When $\mathcal{L}\neq 0$, then,
$N=1,d=4$ supersymmetry implies that the 2-forms $B_{\underline{\rm
    a}},B_{\sharp}$ transform under new St\"uckelberg shifts parametrized by a
2-form gauge transformation parameter $\Lambda$. Still, since $\Lambda \neq 0$
only when $\mathcal{L}\neq 0$, and in this case we have to impose the new
constraint (something we have expressed through Eq.~(\ref{eq:Lconstraint})),
the gauge transformations of the projected 2-forms $Z^{MA}B_{A}$ are left
unchanged by the new 2-form St\"uckelberg shifts. Therefore the field
strengths $F^{M}$ and the action keep their standard form.

In the standard tensor hierarchy it is necessary to introduce 3-forms
$C_{A}{}^{M}$ to construct gauge-covariant field strengths $H_{A}$ for the
2-forms $B_{A}$. These 3-forms are the dual of the embedding tensor
$\vartheta_{M}{}^{A}$.  However, when $\mathcal{L}\neq 0$, the standard tensor
hierarchy field strengths $H_{A}$ need to be modified by the addition of a
3-form $C$, into $H^{\prime}_{A}$, see Eq.~(\ref{eq:HprimeA}). The 3-form $C$
must absorb the new St\"uckelberg shifts of the 2-forms $B_{\underline{\rm
    a}},B_{\sharp}$, but one has to show that $N=1, d=4$ supergravity allows
for such a 3-form.

We have found consistent supersymmetry transformation rules for two 3-forms
$C$ and $C^{\prime}$ the first of which has precisely the required gauge
transformations. $C^{\prime}$ is unexpected from the hierarchy point of view
but turns out to be the dual of the superpotential, seen as a deformation of
the ungauged theory. The fact that it is not predicted by the hierarchy (even
in its extended form which includes the constraint
Eq.~(\ref{eq:superpotentialconstraint3})) is due to the fact that the
superpotential is not associated to any gauge symmetry, which is the basis of
the tensor hierarchy. On the other hand, the existence of the 3-form $C$
suggests the possible existence of another deformation of $N=1,d=4$
supergravity unrelated to gauge symmetry and to the superpotential.

Again, in the $\mathcal{L}\neq 0$ case the field strengths $G_{C}{}^{M}$ need
to be modified by the addition of new 4-forms $D^{M}$ not predicted by the
standard hierarchy, which must absorb gauge transformations related to
$\Lambda$. In the standard hierarchy the 4-forms $D_{E}{}^{NP},D_{AB},D^{NPQ}$
are associated to the constraints $Q_{NP}{}^{E},Q^{AB},L_{NPQ}$. The fourth
4-form that appears when $\mathcal{L}\neq 0$ in $N=1,d=4$ supergravity could
well be related to the constraint $(\vartheta_{M}{}^{\underline{\rm
    a}}\mathcal{P}_{\underline{\rm a}}
+\vartheta_{M}{}^{\sharp}\mathcal{P}_{\sharp})=0$ that the embedding tensor
must satisfy. This can only be fully confirmed by the construction of a
supersymmetric action containing all the $p$-forms as in
\cite{Bergshoeff:2009ph}. Nevertheless, it is clear that, when we vary the
action without any constraints imposed on the embedding tensor, we expect it
to be necessary to introduce a 4-form $D^{M}$ multiplying that constraint. The
gauge transformations of the 4-forms $D^{M}$ should compensate for this lack
of gauge invariance.

Some, but not all, of the $p$-forms in the hierarchy may be associated to
dynamical supersymmetric branes. In order to construct a $\kappa$-symmetric
action for a $(p-1)$-brane that couples to a certain $p$-form, two necessary
conditions are that the $p$-form transforms under no St\"uckelberg shift and
that under supersymmetry transform into a gravitino multiplied by some scalars
may couple to branes. In $N=1, d=4$ supergravity the $p$-forms that satisfy
this condition are the (subset) of 2-forms $B_{\mathbf{a}}$ whose gauge
transformations are massless. These are the 2-forms whose field strengths are
dual to ungauged isometry currents. From the analysis of
\cite{Bergshoeff:2007ij,Ortin:2008wj} we know that these couple to strings
(one-branes that have been referred to as stringy cosmic strings). Another
form which satisfies the criteria is the 3-form $C^{\prime}$ which is a
natural candidate to describe couplings to domain walls. We note that there
are no 1-forms and 4-forms that can couple to a massive brane. There are thus
no 1/2 BPS black holes in the theory and no 1/2 BPS space-time filling
branes. The latter fact may be qualitatively understood from the fact that one
cannot truncate the minimal $N=1, d=4$ supersymmetry algebra to a
supersymmetry algebra with half of the original supercharges.


\section*{Acknowledgments}

M.H.~would like to thank H.~Samtleben for many useful discussions and
the \'Ecole Normale Sup\'erieure de Lyon for its hospitality during the early
stages of this work. JH was supported by a research grant of the Swiss
National Science Foundation as well as by the
``Innovations- und Kooperationsprojekt C-13'' of the Schweizerische
Universit\"atskonferenz SUK/CUS and further wishes to thank the Instituto de F\'isica
Te\'orica of the Universidad Aut\'onoma de Madrid for its hospitality. 
This work has been supported in part by the
INTAS Project 1000008-7928, the Spanish Ministry of Science and Education
grants FPU AP2004-2574 (MH), FPA2006-00783 and PR2007-0073 (TO), the Comunidad
de Madrid grant HEPHACOS P-ESP-00346 and by the Spanish Consolider-Ingenio
2010 program CPAN CSD2007-00042.  Further, TO wishes to express his gratitude
to M.M.~Fern\'andez for her unwavering support.

\appendix

\section{K\"ahler geometry}
\label{app-kahlergeometry}

A K\"ahler manifold is a complex manifold on which there exist complex
coordinates $Z^{i}$ and $Z^{*\, i^{*}} = (Z^{i})^{*}$ and a real function
$\mathcal{K}(Z,Z^{*})$, called the {\em K\"ahler potential}, such that the 

\begin{equation}
ds^{2} = 2 \mathcal{G}_{ii^{*}}\ dZ^{i}dZ^{*\, i^{*}}\, ,
\end{equation}

\noindent
with

\begin{equation}
\label{eq:Kmetric}
\mathcal{G}_{ii^{*}} = \partial_{i}\partial_{i^{*}}\mathcal{K}\, .
\end{equation}

The \textit{K\"ahler (connection) 1-form}  $\mathcal{Q}$ is defined by

\begin{equation}
\label{eq:K1form}
\mathcal{Q} \equiv {\textstyle\frac{1}{2i}}(dZ^{i}\partial_{i}\mathcal{K} -
dZ^{*\, i^{*}}\partial_{i^{*}}\mathcal{K})\, ,
\end{equation}

\noindent
and the \textit{K\"ahler 2-form} $\mathcal{J}$ is its exterior
derivative

\begin{equation}
\label{eq:K2form}
\mathcal{J} \equiv d\mathcal{Q} = i\mathcal{G}_{ii^{*}} 
dZ^{i}\wedge dZ^{*\, i^{*}}\, .
\end{equation}

The K\"ahler potential is defined only up to \textit{K\"ahler transformations}

\begin{equation}
\label{eq:Kpotentialtransformation}
\mathcal{K}^{\prime}(Z,Z^{*})=\mathcal{K}(Z,Z^{*})+f(Z)+f^{*}(Z^{*})\, , 
\end{equation}

\noindent
where $f(Z)$ is any holomorphic function of the complex coordinates $Z^{i}$
that leave the K\"ahler metric and 2-form invariant. The components of the
K\"ahler connection 1-form transform according to

\begin{equation}
\label{eq:K1formtransformation}
\mathcal{Q}^{\prime}_{i} =\mathcal{Q}_{i} 
-{\textstyle\frac{i}{2}}\partial_{i}f\, .
\end{equation}

Objects with K\"ahler weight $(q,\bar{q})$ transform by definition under the
above K\"ahler transformations with a factor $e^{-(qf+\bar{q}f^{*})/2}$ and
their K\"ahler-covariant derivative $\mathcal{D}$ is

\begin{equation}
\label{eq:Kcovariantderivative}
\mathcal{D}_{i} \equiv \nabla_{i} +iq \mathcal{Q}_{i}\, ,
\hspace{1cm}
\mathcal{D}_{i^{*}} \equiv \nabla_{i^{*}} -i\bar{q} \mathcal{Q}_{i^{*}}\, ,
\end{equation}

\noindent
where $\nabla$ is the standard covariant derivative associated to the
Levi-Civit\`a connection. The Ricci identity for this covariant derivative is,
on objects without any indices and K\"ahler weight $(q,\bar{q})$

\begin{equation}
[\mathcal{D}_{i},\mathcal{D}_{j^{*}}] = 
{\textstyle\frac{1}{2}}(\bar{q}-q)\mathcal{G}_{ij^{*}}\, .
\end{equation}

When $(q,\bar{q})=(1,-1)$, this defines a complex line bundle over the
K\"ahler manifold whose first, and only, Chern class equals the K\"ahler
2-form $\mathcal{J}$, i.e.~a \textit{K\"ahler-Hodge (KH) manifold}. These are
the manifolds parametrized by the complex scalars of the chiral multiplets of
$N=1,d=4$ supergravity. Furthermore, objects such as the superpotential and
all the spinors of the theory have a well-defined K\"ahler weight.

We will often use the spacetime pullback of the K\"ahler-covariant derivative
on tensor fields with K\"ahler weight $(q,-q)$ (weight $q$, for short):

\begin{equation}
\label{eq:Kcovariantderivative2}
\mathfrak{D}_{\mu}= \nabla_{\mu} +iq\mathcal{Q}_{\mu}\, ,
\end{equation}

\noindent
where $\nabla_{\mu}$ is the standard spacetime (and/or Lorentz-) covariant
derivative plus possibly the pullback of the Levi-Civit\`a connection.
$\mathcal{Q}_{\mu}$ is the pullback of the K\"ahler 1-form

\begin{equation}
\label{eq:KahlerConPB}
\mathcal{Q}_{\mu} =  
{\textstyle\frac{1}{2i}}(\partial_{\mu}Z^{i}\partial_{i}\mathcal{K} -
\partial_{\mu}Z^{*\, i^{*}}\partial_{i^{*}}\mathcal{K})\, .
\end{equation}


\section{Projectors of the $d=4$ tensor hierarchy}
\label{app-projectors}

The 4-dimensional hierarchy's field strengths are defined in terms of the
invariant tensors $Z^{MA},Y_{AM}{}^{B},W_{C}{}^{MAB},W_{CNPQ}{}^{M},
W_{CNP}{}^{EM}$ which act as projectors. In this appendix we collect  their
definitions and the properties that they satisfy. 

The projectors are defined by 

\begin{eqnarray}
\label{eq:Ztensordef}
Z^{PA} & \equiv & 
-{\textstyle\frac{1}{2}}\Omega^{NP}\vartheta_{N}{}^{A}
=
\left\{
  \begin{array}{l}
    +{\textstyle\frac{1}{2}}\vartheta^{\Lambda A}\, ,\\
\\
    -{\textstyle\frac{1}{2}}\vartheta_{\Lambda}{}^{A}\, ,\\
  \end{array}
\right. \\
& & \nonumber \\
\label{eq:Ytensordef}
Y_{AM}{}^{C} & \equiv & 
\vartheta_{M}{}^{B}f_{AB}{}^{C} -T_{A\, M}{}^{N}\vartheta_{N}{}^{C}\, ,\\
& & \nonumber \\
\label{eq:W1}
W_{C}{}^{MAB} 
& \equiv & 
-Z^{M[A}\delta_{C}{}^{B]}\, , 
\\
& & \nonumber \\
\label{eq:W2}
W_{CNPQ}{}^{M} 
& \equiv & 
T_{C\, (NP}\delta_{Q)}{}^{M}\, ,
\\
& & \nonumber \\
\label{eq:W3}
W_{CNP}{}^{EM} 
& \equiv & 
\vartheta_{N}{}^{D}f_{CD}{}^{E}\delta_{P}{}^{M}
+X_{NP}{}^{M}\delta_{C}{}^{E}
-Y_{CP}{}^{E}\delta_{N}{}^{M}\, .
\end{eqnarray}

They satisfy the orthogonality relations

\begin{eqnarray}
\label{eq:ZY=0}
Z^{MA} Y_{AN}{}^{C} & = & {\textstyle\frac{1}{2}}\Omega^{PM} Q_{PN}{}^{C}= 0\,
,  \\
& & \nonumber \\
Y_{AM}{}^{C}W_{C}{}^{MAB} & = & Y_{AM}{}^{C}W_{CNPQ}{}^{M}
=Y_{AM}{}^{C}W_{CNP}{}^{EM} =0\, .
\end{eqnarray}

The $W$ projectors are related to the embedding tensor constraints by 

\begin{eqnarray}
\label{eq:zW1}
\vartheta_{M}{}^{C} W_{C}{}^{MAB} & = & 2Q^{AB}\, ,\\
& & \nonumber \\
\label{eq:zW2}
\vartheta_{M}{}^{C} W_{CNPQ}{}^{M} & = & L_{NPQ}\, ,\\
& & \nonumber \\
\label{eq:zW3}
\vartheta_{M}{}^{C}W_{CNP}{}^{EM} & = & 2Q_{NP}{}^{E}\, .
\end{eqnarray}

\noindent
Under variations we have

\begin{eqnarray}
\label{eq:dzW1}
\delta \vartheta_{M}{}^{C} W_{C}{}^{MAB} & = & \vartheta_{M}{}^{C} \delta
W_{C}{}^{MAB}  = {\textstyle\frac{1}{2}}\delta (\vartheta_{M}{}^{C}
W_{C}{}^{MAB})=\delta Q^{AB}\, ,\\
& & \nonumber \\
\label{eq:dzW2}
\delta \vartheta_{M}{}^{C} W_{CNPQ}{}^{M} & = & \delta L_{NPQ}\, , \\
& & \nonumber \\
\label{eq:dzW3}
\delta \vartheta_{M}{}^{C} W_{CNP}{}^{EM} & = & 
\vartheta_{M}{}^{C} \delta  W_{CNP}{}^{EM}
={\textstyle\frac{1}{2}} \delta (\vartheta_{M}{}^{C} W_{CNP}{}^{EM}) 
=\delta Q_{NP}{}^{E}\, .
\end{eqnarray}

The constraints
Eqs.~(\ref{eq:quadraticE}), (\ref{eq:quadraticTdef}) and (\ref{eq:linear}) are related through the following identities

\begin{eqnarray}
Q^{AB}Y_{BP}{}^{E}-{\textstyle\frac{1}{2}}Z^{NA}Q_{NP}{}^{E}
& = & 0\, ,
\label{eq:reconstraint1}\\
& & \nonumber \\
Q_{(MN)}{}^{A} -3L_{MNP}Z^{PA} -2Q^{AB}T_{BMN} & = & 0\, ,
\label{eq:reconstraint2}
\end{eqnarray}

\noindent
where Eq.~(\ref{eq:reconstraint1}) can be obtained from Eq.~(\ref{eq:reconstraint2}) by multiplying the latter by $Z^{NE}$. Differentiating these identities with respect to the embedding tensor, using
Eqs.~(\ref{eq:dzW1})-(\ref{eq:dzW3}), we also find the following relations among the $W$
tensors:

\begin{eqnarray}
&&W_{C}{}^{MAB}Y_{BP}{}^{E}-{\textstyle\frac{1}{2}}Z^{NA}W_{CNP}{}^{EM} \nonumber\\
&&-\frac{1}{4}Q^M{}_P{}^E\delta_C^A+Q^{AB}\left[\delta_P^Mf_{BC}{}^E-T_{BP}{}^M\delta_C^E\right] =0\, ,\label{eq:Wrelation1}\\
& & \nonumber \\
&&W_{C(MN)}{}^{AQ} -3W_{CMNP}{}^{Q}Z^{PA} 
-{\textstyle\frac{3}{2}} L_{MN}{}^{Q}\delta_{C}{}^{A} 
-2W_{C}{}^{QAB}T_{B\, MN} =0\, .\label{eq:Wrelation2}
\end{eqnarray}


\section{Gauge transformations and field strengths of the $d=4$ tensor hierarchy}
\label{app-gaugetranshierarchy}

The gauge transformations of the different fields of the tensor hierarchy are

\begin{eqnarray}
\label{eq:deltahAM}
\delta_{h} A^{M} & = & -\mathfrak{D}\Lambda^{M} -Z^{MA}\Lambda_{A}\, ,
\\
& & \nonumber \\
\label{eq:deltahBA}
\delta_{h} B_{A} & = &  \mathfrak{D}\Lambda_{A} +2T_{A\, NP}[\Lambda^{N}F^{P}
+{\textstyle\frac{1}{2}}A^{N}\wedge\delta_{h} A^{P}] 
-Y_{AM}{}^{C}\Lambda_{C}{}^{M}\, ,
\\
& & \nonumber \\
\label{eq:deltahCCM}
\delta_{h} C_{C}{}^{M} 
& = & 
\mathfrak{D}\Lambda_{C}{}^{M}
 -F^{M}\wedge\Lambda_{C}
 -\delta_{h} A^{M}\wedge B_{C} 
-{\textstyle\frac{1}{3}}T_{C\, NP} A^{M}\wedge A^{N} \wedge \delta_{h} A^{P}
\nonumber \\
& & \nonumber \\
& & 
+\Lambda^{M}H_{C}
-W_{C}{}^{MAB}\Lambda_{AB} -W_{CNPQ}{}^{M}\Lambda^{NPQ}
-W_{CNP}{}^{EM}\Lambda_{E}{}^{NP}\, ,
\\
& & \nonumber \\
\label{eq:deltahDAB}
\delta_{h} D_{AB} & = & \mathfrak{D}\Lambda_{AB}
+2 T_{[AMN}\tilde{\Lambda}_{B]}{}^{(MN)}
+Y_{[A|P}{}^{E}(\Lambda_{B]E}{}^{P}-B_{B]}\wedge \Lambda_{E}{}^{P})
+\mathfrak{D}\Lambda_{[A}\wedge B_{B]}
\nonumber \\
& & \nonumber \\
& & 
-2\Lambda_{[A}\wedge H_{B]} 
+2T_{[A| NP}[\Lambda^{N}F^{P}
-{\textstyle\frac{1}{2}}A^{N}\wedge \delta_{h} A^{P}]\wedge B_{|B]}\, , \\
& & \nonumber \\
\label{eq:deltahDENP}
\delta_{h} D_{E}{}^{NP} & = & \mathfrak{D}\Lambda_{E}{}^{NP} 
+\tilde{\Lambda}_{E}{}^{(NP)}
+{\textstyle\frac{1}{2}}Z^{NB}\Lambda_{BE}{}^{P} -F^{N}\wedge \Lambda_{E}{}^{P}
\nonumber \\
& & \nonumber \\
& &
+C_{E}{}^{P}\wedge \delta_{h} A^{N} 
+{\textstyle\frac{1}{12}} T_{EQR} A^{N}\wedge A^{P} \wedge A^{Q} 
\wedge \delta_{h} A^{R}
+\Lambda^{N}G_{E}{}^{P}\, ,\\
& & \nonumber \\
\label{eq:deltahDNPQ}
\delta_{h} D^{NPQ} & = & \mathfrak{D}\Lambda^{NPQ}
-3 Z^{(N|A}\tilde{\Lambda}_{A}{}^{|PQ)}
-2A^{(N}\wedge dA^{P}\wedge \delta_{h} A^{Q)}
 \nonumber \\
& & \nonumber \\
& &
-{\textstyle\frac{3}{4}} X_{RS}{}^{(N} 
A^{P|} \wedge A^{R}\wedge A^{S} \wedge  \delta_{h} A^{|Q)}
-3\Lambda^{(N}F^{P}\wedge F^{Q)}
\, ,
\end{eqnarray}

\noindent
where we remark that $\Lambda_{E}{}^{NP}$ is a 3-form and
$\tilde{\Lambda}_{E}{}^{(NP)}$ is a 4-form.

Their gauge-covariant field strengths are

\begin{eqnarray}
F^{M} 
& = & 
dA^{M} +{\textstyle\frac{1}{2}}X_{[NP]}{}^{M}A^{N}\wedge A^{P} +Z^{MA}B_{A}\, ,  
\\
& & \nonumber \\
H_{A} & = & \mathfrak{D}B_{A}
+T_{A\, RS}A^{R}\wedge[dA^{S} +{\textstyle\frac{1}{3}}X_{NP}{}^{S}A^{N}\wedge
A^{P}] +Y_{AM}{}^{C}C_{C}{}^{M}\, ,
\label{eq:HA}
\\
& & \nonumber \\
G_{C}{}^{M} 
& = &  
\mathfrak{D}C_{C}{}^{M} +[F^{M}-{\textstyle\frac{1}{2}}Z^{MA}B_{A}]\wedge B_{C}
+{\textstyle\frac{1}{3}}T_{C\, SQ} A^{M}\wedge A^{S} \wedge dA^{Q}
\nonumber\\
& & \nonumber \\
& & 
+{\textstyle\frac{1}{12}}T_{C\, SQ}X_{NT}{}^{Q}A^{M}\wedge A^{S} \wedge A^{N}
\wedge A^{T}
\nonumber \\
& & \nonumber \\
& & 
+W_{C}{}^{MAB}D_{AB} +W_{CNPQ}{}^{M}D^{NPQ}
+W_{CNP}{}^{EM} D_{E}{}^{NP}\, .
\label{eq:GCM}
\end{eqnarray}

These field strengths are related by the following hierarchical Bianchi
identities

\begin{eqnarray}
\mathfrak{D}F^{M} 
& = & 
Z^{MA}H_{A}\, ,
\\
& & \nonumber \\  
\mathfrak{D}H_{A} 
& = & 
Y_{AM}{}^{C} G_{C}{}^{M}  +T_{A\, MN}F^{M}\wedge F^{N}\, .
\end{eqnarray}



\begin{thebibliography}{99}

\bibitem{Trigiante:2007ki}
M.~Trigiante,
\hepth{0701218}.

\bibitem{Weidner:2006rp}
M.~Weidner,
Fortsch.\ Phys.\  {\bf 55} (2007) 843
[\hepth{0702084}].

\bibitem{Samtleben:2008pe}
H.~Samtleben,
Class.\ Quant.\ Grav.\  {\bf 25} (2008) 214002
[\arxiv{0808.4076} [hep-th]].

\bibitem{Cordaro:1998tx}
F.~Cordaro, P.~Fr\'e, L.~Gualtieri, P.~Termonia and M.~Trigiante,
Nucl.\ Phys.\  B {\bf 532} (1998) 245
[\hepth{9804056}].

\bibitem{deWit:2002vt}
B.~de Wit, H.~Samtleben and M.~Trigiante,
Nucl.\ Phys.\  B {\bf 655} (2003) 93
[\hepth{0212239}].

\bibitem{deWit:2003hq}
B.~de Wit, H.~Samtleben and M.~Trigiante,
Phys.\ Lett.\  B {\bf 583} (2004) 338
[\hepth{0311224}].

\bibitem{deWit:2005hv}
B.~de Wit and H.~Samtleben,
Fortsch.\ Phys.\  {\bf 53} (2005) 442
[\hepth{0501243}].

\bibitem{deWit:2005ub}
B.~de Wit, H.~Samtleben and M.~Trigiante,
JHEP {\bf 0509} (2005) 016
[\hepth{0507289}].



\bibitem{deWit:2004nw}
B.~de Wit, H.~Samtleben and M.~Trigiante,
Nucl.\ Phys.\  B {\bf 716} (2005) 215
[\hepth{0412173}].

\bibitem{Samtleben:2005bp}
H.~Samtleben and M.~Weidner,
Nucl.\ Phys.\  B {\bf 725} (2005) 383
[\hepth{0506237}].

\bibitem{Schon:2006kz}
J.~Schon and M.~Weidner,
JHEP {\bf 0605} (2006) 034
[\hepth{0602024}].

\bibitem{de Wit:2007mt}
B.~de Wit, H.~Samtleben and M.~Trigiante,
JHEP {\bf 0706} (2007) 049
[\arxiv{0705.2101} [hep-th]].

\bibitem{Bergshoeff:2007vb}
E.~A.~Bergshoeff, J.~Gomis, T.~A.~Nutma and D.~Roest,
JHEP {\bf 0802} (2008) 069
[\arxiv{0711.2035} [hep-th]].

\bibitem{deWit:2008ta}
B.~de Wit, H.~Nicolai and H.~Samtleben,
JHEP {\bf 0802} (2008) 044
[\arxiv{0801.1294} [hep-th]].


\bibitem{Bergshoeff:2009ph}
E.~A.~Bergshoeff, J.~Hartong, O.~Hohm, M.~H\"ubscher and T.~Ort\'{\i}n,
\arxiv{0901.2054} [hep-th].


\bibitem{deWit:2009zv}
B.~de Wit and M.~van Zalk,
Gen.\ Rel.\ Grav.\  {\bf 41} (2009) 757
[arXiv:0901.4519 [hep-th]].


\bibitem{kn:toappear2}
M.~H\"ubscher, P.~Meessen, T.~Ort\'{\i}n and Silvia  Vaul\`a,
``All the supersmmetric solutions of gauged, matter-coupled,
 $N=1$ and $N=2$ $d=4$ supergravity'' 
(in preparation).

\bibitem{kn:toappear3}
J.~Hartong, T.~Ort\'{\i}n  and Silvia  Vaul\`a,
``The general $d=5$ and $d=6$ tensor hierarchies'' 
(in preparation).

\bibitem{Ortin:2008wj}
T.~Ort\'{\i}n,
JHEP {\bf 0805} (2008) 034
[\arxiv{0802.1799} [hep-th]].


\bibitem{Gaillard:1981rj}
M.~K.~Gaillard and B.~Zumino,
Nucl.\ Phys.\  B {\bf 193} (1981) 221.

\bibitem{DeRydt:2008hw}
J.~De Rydt, T.~T.~Schmidt, M.~Trigiante, A.~Van Proeyen and M.~Zagermann,
\arxiv{0808.2130} [hep-th].



\bibitem{Bergshoeff:2001pv}
E.~Bergshoeff, R.~Kallosh, T.~Ort\'{\i}n, D.~Roest and A.~Van Proeyen,
Class.\ Quant.\ Grav.\  {\bf 18} (2001) 3359
[\hepth{0103233}].

\bibitem{Bergshoeff:2005ac}
E.~A.~Bergshoeff, M.~de Roo, S.~F.~Kerstan and F.~Riccioni,
JHEP {\bf 0508} (2005) 098
[\hepth{0506013}].

\bibitem{Bergshoeff:2006qw}
E.~A.~Bergshoeff, M.~de Roo, S.~F.~Kerstan, T.~Ortin and F.~Riccioni,
JHEP {\bf 0607} (2006) 018
[\hepth{0602280}].

\bibitem{Bergshoeff:2006gs}
E.~A.~Bergshoeff, M.~de Roo, S.~F.~Kerstan, T.~Ortin and F.~Riccioni,
JHEP {\bf 0702} (2007) 007
[\hepth{0611036}].

\bibitem{Gomis:2007gb}
J.~Gomis and D.~Roest,
JHEP {\bf 0711} (2007) 038
[\arxiv{0706.0667} [hep-th]].

\bibitem{Bergshoeff:2007ij}
E.~A.~Bergshoeff, J.~Hartong, M.~H\"ubscher and T.~Ort\'{\i}n,
JHEP {\bf 0805} (2008) 033
[\arxiv{0711.0857} [hep-th]].

\bibitem{Cremmer:1982en}
E.~Cremmer, S.~Ferrara, L.~Girardello and A.~Van Proeyen,
Nucl.\ Phys.\  B {\bf 212} (1983) 413.

\bibitem{de Wit:1984px}
B.~de Wit, P.~G.~Lauwers and A.~Van Proeyen,
Nucl.\ Phys.\  B {\bf 255} (1985) 569.

\bibitem{Andrianopoli:2004sv}
L.~Andrianopoli, S.~Ferrara and M.~A.~Lledo,
JHEP {\bf 0404} (2004) 005
[\hepth{0402142}].

\bibitem{de Vroome:2007zd}
M.~de Vroome and B.~de Wit,
JHEP {\bf 0708} (2007) 064
[\arxiv{0707.2717} [hep-th]].



\end{thebibliography}
\end{document}